\documentclass{aa}  

\usepackage{graphicx}
\usepackage{txfonts}
\usepackage{lipsum}
\usepackage{subcaption}         
\usepackage{lscape}             
\usepackage{placeins}           
                                
\usepackage{color}
\usepackage{natbib,twoopt}
\usepackage[hyphenbreaks]{breakurl}
\usepackage[breaklinks]{hyperref}
\hypersetup{colorlinks=true, linkcolor=blue, citecolor=blue, filecolor=green, urlcolor=blue }

\newcommand{\pcm}{\,cm$^{-2}$}	
\newcommand{\psec}{s$^{-1}$}  
\newcommand{\erg}{erg cm$^{-2}$ s$^{-1}$} 
\newcommand{\lum}{erg s$^{-1}$} 
\newcommand{\lxp}{\emph{AstroSat}/LAXPC}

\def \swift {\emph{Swift}}

\def \src{{RX J0520.5--6932}}
\def \swiftxrt{\emph{Swift}-XRT}

\def \nustar{\emph{NuSTAR}}

\def \nicer{\emph{NICER}}

\def \maxi{\emph{MAXI}}

\def \astrosat{\emph{AstroSat}}

\begin{document}

   \title{Probing accretion dynamics and spin evolution in the X-ray pulsar \src\ during its 2024 outburst}
    \titlerunning{Accretion dynamics and spin evolution of \src}


   \author{Rahul Sharma\inst{1,2}
        \and Aru Beri\inst{3,4,5}
        \and Biswajit Paul\inst{2}
        \and Andrea Sanna\inst{6}
        \and Chandreyee Maitra\inst{1,7}
        \and Haonan Yang\inst{7,8,9}
        }

   \institute{Inter-University Centre for Astronomy and Astrophysics (IUCAA), Ganeshkhind, Pune 411007, India\\
    \email{rahul1607kumar@gmail.com}
        \and Raman Research Institute, C V Raman Avenue, Sadashivanagar, Bangalore 560080, India
        \and Indian Institute of Science Education and Research (IISER) Mohali, Punjab 140306, India
        \and School of Physics \& Astronomy, University of Southampton, Southampton, Hampshire SO17 1BJ, UK
        \and Indian Institute of Astrophysics, Koramangala II Block, Bangalore-560034, India
        \and Universit\'{a} degli Studi di Cagliari, Dipartimento di Fisica, SP Monserrato-Sestu, KM 0.7, 09042 Monserrato, Italy
        \and Max-Planck-Institut für extraterrestrische Physik, Gießenbachstraße 1, D-85748 Garching bei München, Germany
        \and National Astronomical Observatories, Chinese Academy of Sciences, 20A Datun Road, Beijing 100101, China
        \and School of Astronomy and Space Science, University of Chinese Academy of Sciences, 19A Yuquan Road, Beijing 100049, China
             }
    

 
  \abstract
   {After nearly a decade of quiescence, the transient Be/X-ray binary pulsar \src\ underwent an outburst in 2024. We performed X-ray monitoring of the source with \nicer\ and \astrosat\ near the peak of the event.}
   {Our primary objective is to investigate the energy and luminosity dependence of the pulsed emission, characterize the spin evolution, and study the broadband X-ray spectral properties of \src\ during the outburst.}
   {We extracted light curves and spectra from \nicer\ and \astrosat\ observations carried out during the outburst. Pulsations were detected using epoch-folding techniques, enabling a detailed study of pulse-profile evolution as a function of energy and intensity. Broadband spectral modelling was performed using simultaneous data from SXT, LAXPC, and \nicer. The spectra from individual \nicer\ observations were used to study spectral variability.}
   {The \lxp\ and \nicer\ light curves reveal pronounced short-duration flaring activity lasting $\sim$400--700 s, with intensity enhancements by a factor of $\sim$2. The pulse profile exhibits a strong dependence on both energy and intensity, evolving from a simple single-peaked structure at low energies to complex multi-peaked shapes at intermediate energies, and reverting to simpler morphologies at higher energies. Pulse profiles during the flares differ significantly from those in the persistent state, indicating changes in the pulsed beam pattern with a change in the intensity on a short timescale. Broadband spectral analysis reveals a soft excess and an emission feature at $\sim$1 keV, likely arising from reprocessed emission in the accretion disc and fluorescence from Ne K and Fe L ions. Continuous \nicer\ monitoring over nearly one orbital cycle enabled us to track spin evolution with accretion-driven spin-up and spectral variability in the soft X-ray band. Additionally, a declining spin-up rate is observed during the outburst, likely due to a gradual reduction in mass accretion rate.}
   {Our results provide a comprehensive view of the complex accretion dynamics in \src\ during its 2024 outburst. The strong variability in pulse shape and spin behaviour highlights rapid changes in the accretion geometry and torque as a function of accretion rate.}


   \maketitle

\nolinenumbers

\section{Introduction}
\label{intro}

Accretion-powered X-ray pulsars (XRPs) are magnetized neutron stars in binary systems, where accretion of matter from a companion star gives rise to pulsed X-ray emission. A majority of XRPs are found in high-mass X-ray binaries \citep[HMXBs;][]{Liu06}, which typically host either a Be-type star or an OB supergiant companion. Among these, systems composed of a Be star and a neutron star, often in an eccentric orbit, are classified as Be/X-ray binaries \citep[BeXRBs;][]{Reig11}. These systems are often transient in nature and are primarily detected during outburst phases, either regular Type-I outbursts, which occur near periastron as the neutron star passes through the circumstellar disc of the Be star, or more energetic and less frequent Type-II outbursts, which are likely triggered by large-scale mass ejections from the Be star. The latter can occur at any orbital phase and are sometimes linked to warping or structural changes in the Be disc. Due to their strong variability and wide dynamic range, transient X-ray pulsars serve as valuable laboratories for studying accretion physics under different luminosity regimes.

\src\ is a BeXRB system located in the Large Magellanic Cloud (LMC), initially discovered through observations by \emph{ROSAT} \citep{Schmidtke94}. A major outburst was recorded in 1995, with concurrent detection in optical and X-ray bands \citep{Edge04}. During a giant outburst in 2014, the source's X-ray luminosity reached values close to the Eddington limit for a neutron star \citep{Vasilopoulos14a}. Coherent X-ray pulsations at a period of $\sim$8 s confirmed its nature as an accreting X-ray pulsar \citep{Vasilopoulos14b}. Notably, \citet{Tendulkar14} reported the detection of a cyclotron resonant scattering feature (CRSF) at $\sim$31--32 keV, corresponding to a surface magnetic field strength of the order of $\sim10^{12}$ G.

After nearly a decade of quiescence, \src\ re-entered an outburst phase in late March 2024, as reported by multiple observatories \citep{Semena24, Sharma24atel, Zhang24}. The event prompted extensive follow-up observations using multiple instruments such as \astrosat, Neutron star Interior Composition ExploreR (\nicer), Neil Gehrels Swift Observatory (\swift), Nuclear Spectroscopic Telescope ARray (\nustar), the Einstein Probe (EP), and Lobster Eye Imager for Astronomy (LEIA) \citep[e.g,][]{Yang25}. While the orbital parameters of the system have been previously studied \citep{Kuehnel14, Karaferias23}, inconsistencies remain among reported values, particularly between those derived from X-ray and optical data \citep{Vasilopoulos14a}. Notably, a degeneracy in the orbital period between 23.91 and 24.38 days has been reported based on combined spin frequency evolution during the 2014 and 2024 outbursts \citep{Yang25}.

In this work, we present the results from a detailed timing and spectral study of \src\ using a Target of Opportunity (ToO) observation with \astrosat\ and \nicer\ observations obtained during the peak of its 2024 outburst. The broadband spectral properties were examined by combining data from the Soft X-ray Telescope (SXT) and Large Area X-ray Proportional Counter (LAXPC) onboard \astrosat\ and simultaneous \nicer\ observation. The structure of the paper is as follows: Section \ref{obs} describes the observations and data reduction procedures. Section \ref{res} presents the results of the timing and spectral analysis. We discuss our findings in Section \ref{dis} and conclude in Section \ref{summary}. 


\begin{figure}
\centering
 \includegraphics[width=0.9\linewidth]{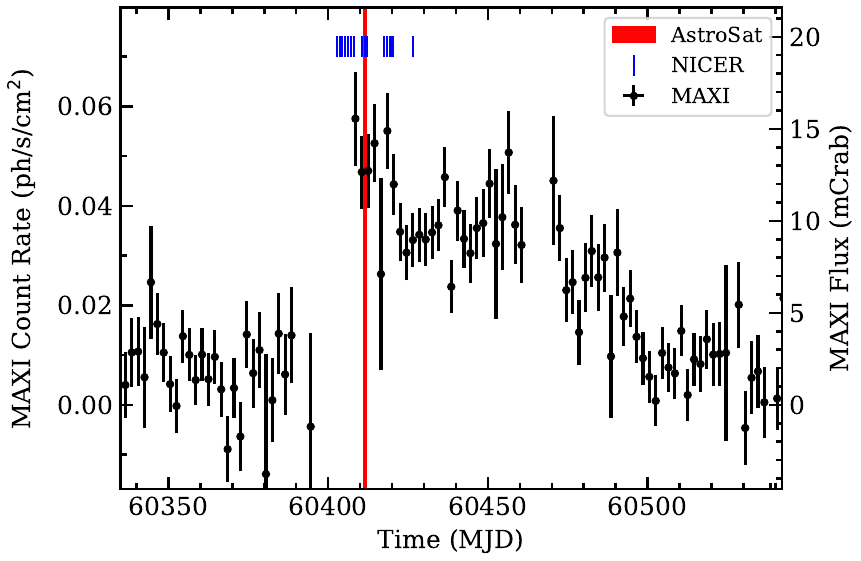}
 \caption{The 2--20 keV light curve of \src\ during its 2024 outburst from \emph{MAXI}-GSC binned at two days. The \emph{MAXI} data points represent the source intensity, whereas the solid red vertical line marks the epoch of \astrosat{} observation, and the blue tick markers indicate the epoch of \nicer\ pointings.}
\label{fig:maxi}
\end{figure}

\begin{table}
    \centering
    \caption{The log of X-ray observations of \src\ analysed in this work.}
    \resizebox{\linewidth}{!}{
    \begin{tabular}{cccccccc}
    \hline
Instrument & Obs-ID    & \multicolumn{2}{c}{Obs Start Date}  & Exposure\\
 &       & (yy-mm-dd) & (MJD)   &  (s)\\
\hline
\astrosat-LAXPC &  9000006180 & 2024-04-11 & 60411.1895 & 53572 \\ 
\astrosat-SXT &  9000006180 & 2024-04-11 & 60411.2268 & 16521 \\[1ex] 

\nicer	&	7204300101	&	2024-04-02	&	60402.8997	&	795	\\
\nicer	&	7204300102	&	2024-04-03	&	60403.8682 &	666	\\
\nicer	&	7204300103	&	2024-04-04	&	60404.3235	&	2310	\\
\nicer	&	7204300104	&	2024-04-05	&	60405.3536 &	2157 \\
\nicer	&	7204300105	&	2024-04-06	&	60406.2583	&	1644	\\
\nicer	&	7204300106	&	2024-04-07	&	60407.2923	&	1337	\\
\nicer	&	7204300107	&	2024-04-08	&	60408.0692	&	1985	\\
\nicer	&	7204300109	&	2024-04-10	&	60410.5734 &	512	\\
\nicer	&	7204300110	&	2024-04-11	&	60411.6064	&	1357	\\
\nicer	&	7204300111	&	2024-04-12	&	60412.2568 &	837	\\
\nicer	&	7204300114	&	2024-04-17	&	60417.3523 &	851	\\
\nicer	&	7204300115	&	2024-04-18	&	60418.3893 &	1569	\\
\nicer	&	7204300116	&	2024-04-19	&	60419.3586	&	322	\\
\nicer	&	7204300117	&	2024-04-20	&	60420.1284 &	497	\\
\nicer	&	7204300119	&	2024-04-26	&	60426.6424 &	379	\\
    \hline     
    \end{tabular}}    
    \label{tab:obs}
\end{table}

\begin{figure}
\centering
 \includegraphics[width=0.9\linewidth]{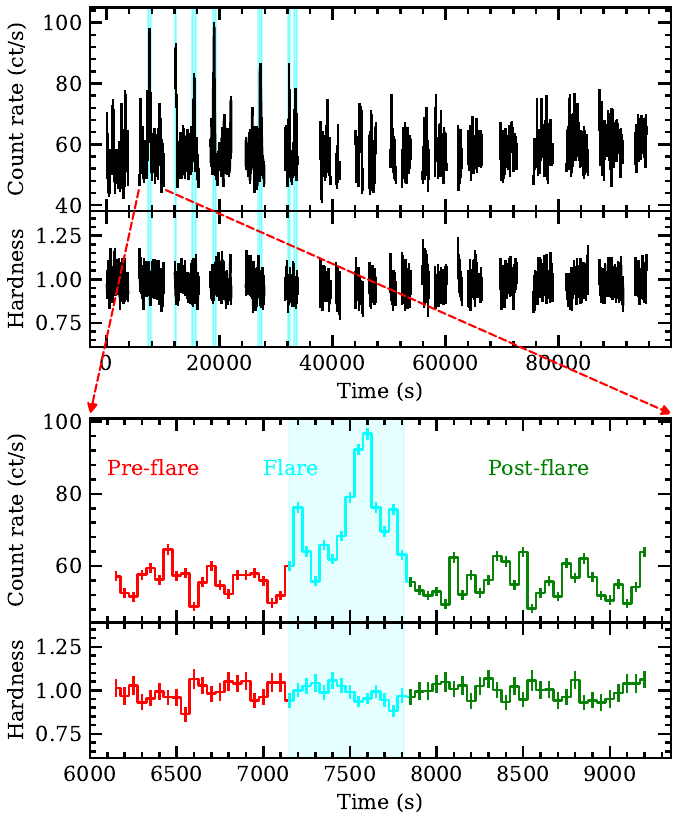}
 \caption{The \lxp\ light curve of \src\ during its 2024 outburst in the 3--25 keV energy range, binned at 50 seconds. The bottom subplot highlights a zoomed-in segment of the light curve (from the second orbit) showing short-timescale flaring events, where different colors (red, cyan, and green) mark pre-flare, flare, and post-flare intervals, respectively. The bottom panel of each subplot displays the hardness ratio, defined as the ratio of count rates in the 8--25 keV and 3--8 keV energy bands. The shaded region highlights the detected flaring events.}
\label{fig:lc}
\end{figure}


\section{Observation and data analysis}
\label{obs}

Figure \ref{fig:maxi} presents the outburst light curve of \src\ during the 2024 outburst with the Gas Slit Camera (GSC) onboard the Monitor of All-sky X-ray Imager \citep[\maxi;][]{Matsuoka09}. The \astrosat\ observation was carried out on 2024 April 11 (MJD 60411), covering nearly one day, marked as the red-shaded vertical region in Fig. \ref{fig:maxi}. Meanwhile, \nicer\ monitored the source from 2024 April 2 to 26 (MJD 60402--60426), around the peak of the outburst, indicated by vertical blue tick marks. Observation details are provided in Table \ref{tab:obs}.

\subsection{\astrosat}

\astrosat\ is India's first dedicated multi-wavelength astronomy satellite \citep{Agrawal2006, Singh2014}, launched in 2015. In this work, we analyze data from SXT and LAXPC. 

\subsubsection{LAXPC}

LAXPC is one of the primary instruments aboard \astrosat. It consists of three co-aligned identical proportional counters (LAXPC10, LAXPC20, and LAXPC30) that work in the energy range of 3--80 keV. Each LAXPC detector independently records the arrival time of each photon with a time resolution of $10~\mu$s and has five layers \citep[for details see][]{Yadav2016, Antia2017}.

During our observation, LAXPC10 was operating at low gain, and LAXPC30 was offline. Therefore, we used data only from the LAXPC20 detector for our analysis.  We utilized the Event Analysis (EA) mode data and processed it using \textsc{LaxpcSoft}\footnote{\url{http://www.tifr.res.in/~astrosat\_laxpc/LaxpcSoft.html}} version 3.4.4 software package to extract light curves. Background estimation was performed using blank sky observations as described in \citet{Antia2017}, and appropriate response files were used for energy calibration. The source and background spectra were extracted using the faint source model following the method outlined in \citet{Misra21}.

We corrected the LAXPC photon arrival times to the Solar system barycentre using the \textsc{as1bary}\footnote{\url{http://astrosat-ssc.iucaa.in/?q=data\_and\_analysis}} tool with the JPL-DE405 ephemeris. We used the best available position of the source, R.A. (J2000)$=05^h 20^m 30.90^s$ and Dec. (J2000) $=-69^{\circ} 31' 55.0''$ \citep{Bonanos09}.

\subsubsection{SXT}

SXT is a focusing X-ray telescope with CCD in the focal plane that can perform X-ray imaging and spectroscopy in the 0.3--7 keV energy range \citep{Singh2016, Singh2017}. \src\ was observed in the Photon Counting (PC) mode with SXT. Level 1 data were processed with \texttt{AS1SXTLevel2-1.4b} pipeline software to generate level 2 cleaned event files. These cleaned files from individual orbits were merged using the SXT event merger tool\footnote{\url{https://github.com/gulabd/SXTMerger.jl}}. The merged event file was then used to extract images, light curves, and spectra using the \textsc{xselect} task, provided as part of \textsc{heasoft} version 6.31.1. A circular region with a radius of 16 arcmin centred on the source was used. No source pile-up was observed as the count rate was below the threshold limit of pileup ($<$40 counts \psec) in the PC mode\footnote{\url{https://www.tifr.res.in/~astrosat_sxt/instrument.html}}. For spectral analysis, we have used the blank sky SXT spectrum as background (SkyBkg\_sxt\_LE0p35\_R16p0\_v05\_Gd0to12.pha) and spectral redistribution matrix file (sxt\_pc\_mat\_g0to12.rmf) provided by the SXT team\footnote{\url{http://www.tifr.res.in/~astrosat\_sxt/dataanalysis.html}}. We generated the correct off-axis auxiliary response files (ARF) using the sxtARFModule tool from the on-axis ARF (sxt\_pc\_excl00\_v04\_20190608.arf) provided by the SXT instrument team. The SXT spectrum was optimally rebinned using \textsc{ftgrouppha} to have a minimum of 25 counts per bin \citep{Kaastra}. Owing to a low duty cycle ($\sim$20\%) and a time
resolution of $\sim$2.3 s, we have not used SXT data for timing analysis in this work.

\subsection{NICER}

\nicer\ \citep{nicer} is a soft X-ray telescope installed on the International Space Station (ISS) in June 2017, equipped with the X-ray Timing Instrument (XTI). The XTI consists of 56 co-aligned Focal Plane Modules (FPMs), each made up of an X-ray concentrator optic associated with a silicon drift detector. The instrument provides high temporal resolution of $\sim$100 ns, high spectral resolution of $\sim$85 eV at 1 keV, and a large effective area of $\sim$1900 cm$^2$ at 1.5 keV, utilizing 52 active detectors. Each \nicer\ observation typically comprises multiple short-duration pointings of the source called snapshots, primarily driven by ISS orbit and visibility constraints. 

\nicer{} monitored \src\ from MJD 60402 to 60426 (Table \ref{tab:obs}). The data were processed using \textsc{heasoft} version 6.33.2 and the \nicer\ Data Analysis Software (\texttt{nicerdas}) version 2024-02-09\_V012A with Calibration Database (CALDB) version xti20240206. Standard calibration and screening criteria such as \emph{cor\_range=1.5--*, max\_lowmem=250} and \emph{threshfilter=NIGHT} were applied using \texttt{nicerl2}.  We extracted spectral files and responses (ARF and RMF files) using \texttt{nicerl3-spect}. The background was obtained using the 3C50 model \citep{Remillard22}. The energy band of \nicer\ analysis was limited to 0.5--10 keV. Light curves were extracted using \texttt{xselect}. The event times were corrected to solar system barycentre using \texttt{barycorr} with the JPL-DE405 ephemeris. 

\begin{figure}
\centering
 \includegraphics[width=\linewidth]{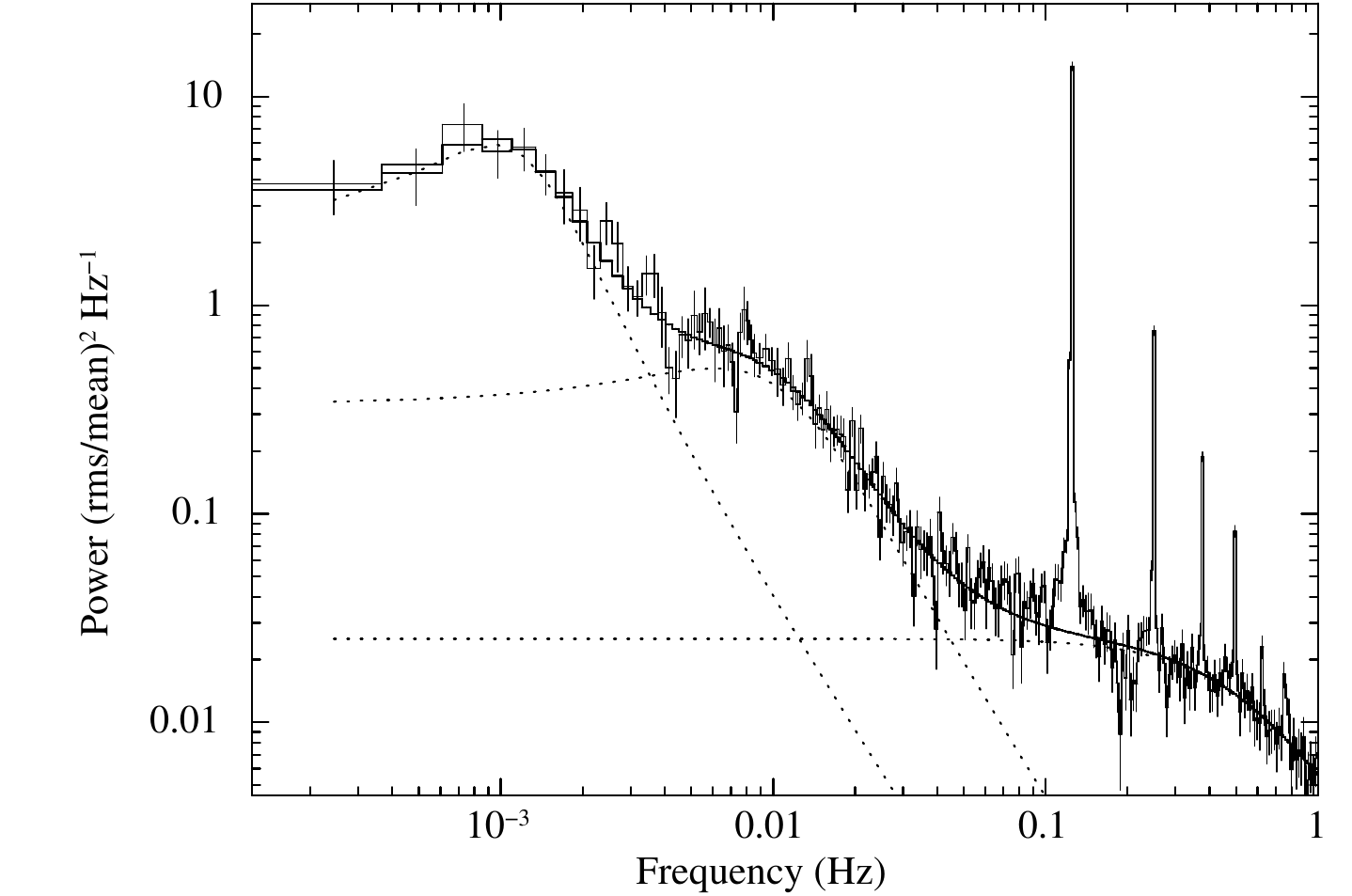}
\caption{Power density spectrum of \src\ in the 3--25 keV band obtained from \lxp\ data. The continuum was modeled using a combination of three Lorentzian components representing broad noise features at different characteristic frequencies. Sharp peaks due to the neutron star's spin frequency and its harmonics were excluded from the fit.}
\label{fig:pds}
\end{figure}

\begin{table}
    \centering
    \caption{Spin ephemeris parameters derived from the \lxp\ observation, along with the evolution of the spin frequency during the 2024 outburst, based on the measurements from \nicer, \astrosat, and \nustar\ observations. All errors reported in this table are at 68\% (1$\sigma$) confidence level.}
    \resizebox{\linewidth}{!}{
    \begin{tabular}{ccccc}
    \hline
Parameters & \astrosat & \multicolumn{2}{c}{Spin evolution}\\
& & Linear \tablefootmark{a} & Quadratic \\
\hline
$T_0$ (MJD) & & 60411.0 & \\
$\nu_0$ (mHz) & 124.56286 (15) & 124.56266 (14) &  124.56298 (12) \\
$\dot{\nu}$ ($10^{-11}$ Hz \psec) & 2.4 (2) & 2.45 (8) & 2.10 (4)  \\
$\ddot{\nu}$ ($10^{-17}$ Hz s$^{-2}$) & - & - & $-$2.11 (14) \\
\hline
    \end{tabular}}  
    \tablefoot{\tablefoottext{a}{The last measurement of \nicer\ at MJD 60418 was not included in this fit.}}
    \label{tab:spin}
\end{table}

\begin{figure}
\centering
 \includegraphics[width=\linewidth]{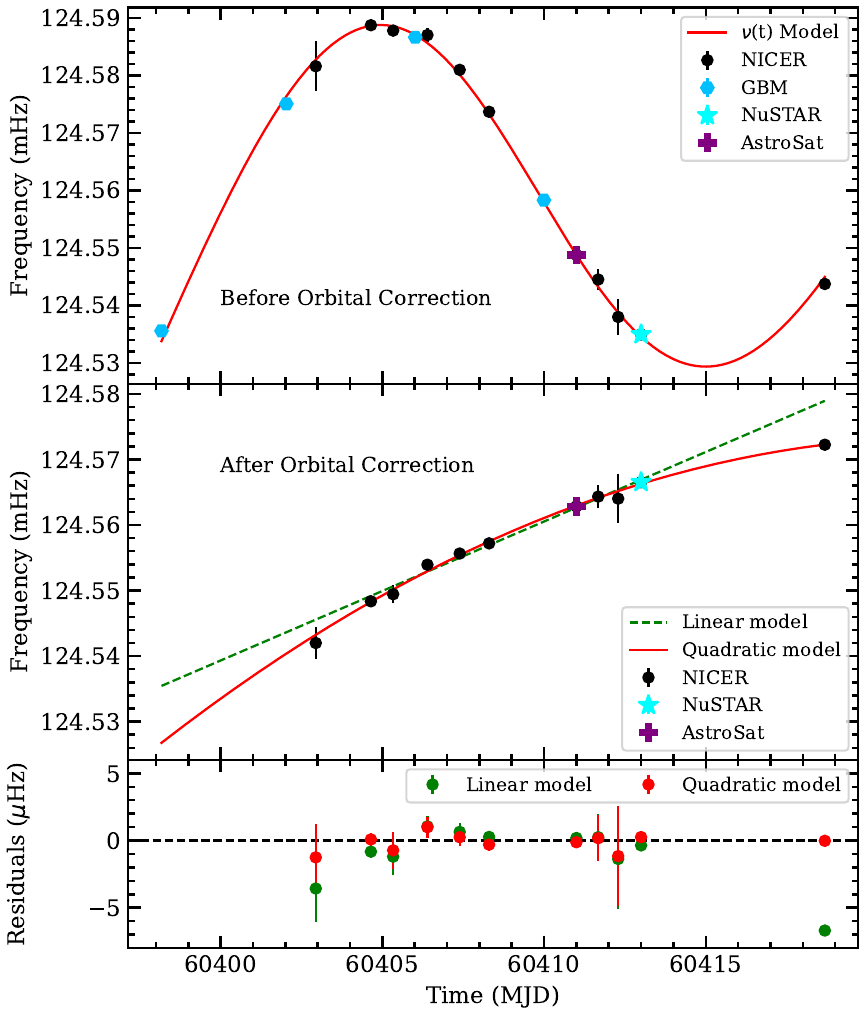}
 \caption{Spin frequency evolution of \src\ during the 2024 outburst. 
 \textit{Top:} Barycentre-corrected spin frequencies from \nicer, \astrosat, \emph{Fermi}/GBM and \nustar, overlaid with the composite model $\nu(t) = \nu_{\rm int}(t) - \nu_{\rm orb}(t)$ (solid red line), which combines the intrinsic quadratic spin evolution with the fixed orbital parameters from \citet{Yang25} (Solution II).
 \textit{Middle:} Spin frequencies after correcting for orbital motion. A quadratic model (solid line) fits all data points and captures the evolving spin-up trend, while a linear model (dashed line) fits only the early part of the outburst (excluding the final \nicer\ point).
 \textit{Bottom:} Residuals of the quadratic and linear fits. The residuals clearly favor the quadratic model, supporting the presence of a varying spin-up rate during the outburst.}
\label{fig:spin}
\end{figure}

\begin{figure}
\centering
 \includegraphics[width=\linewidth]{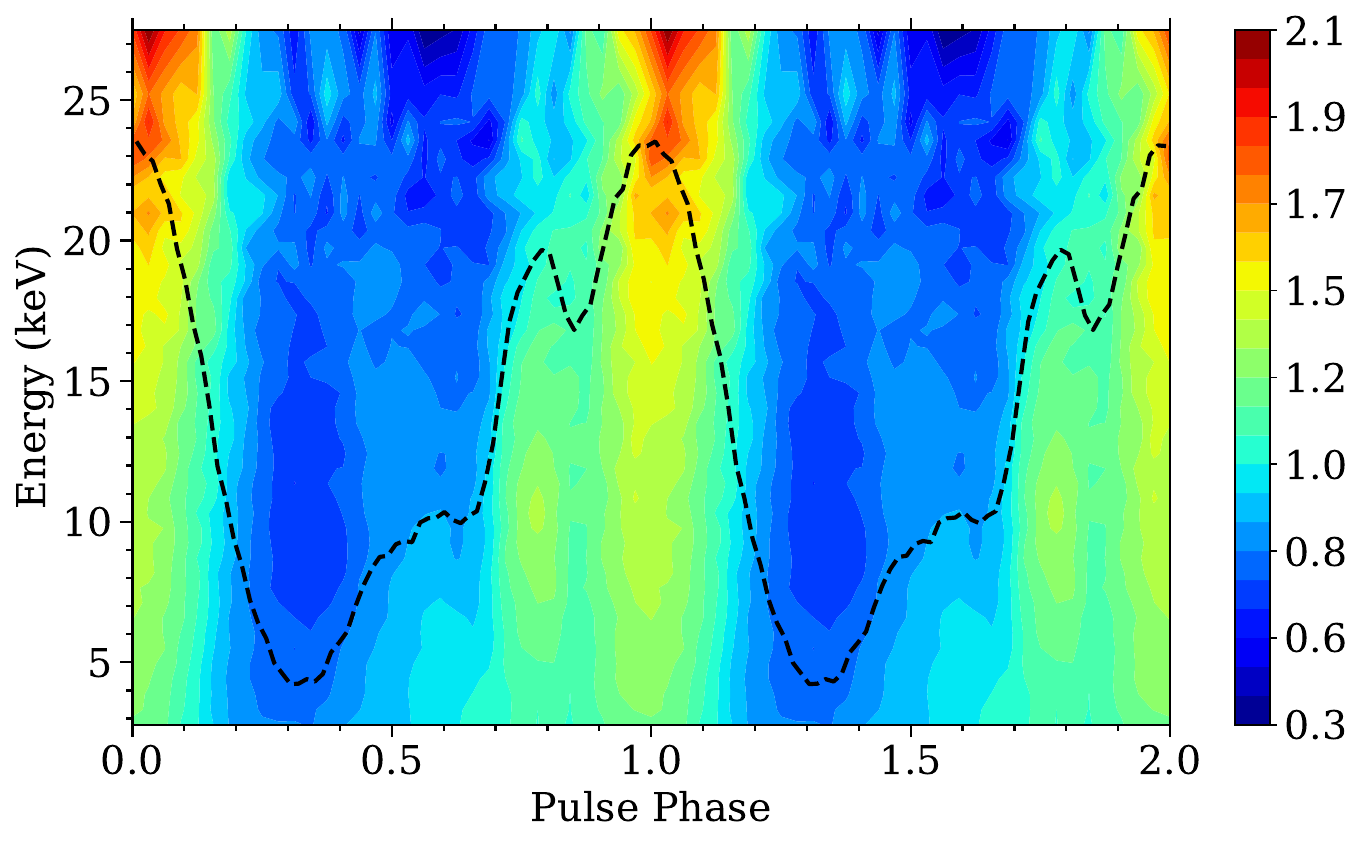}
 \caption{Energy-resolved dynamic pulse profiles of \src{} obtained from \lxp\ data during the 2024 outburst. The colormap represents the normalized pulse intensity across different energy bands. Strong energy dependence is evident, with the secondary peak most prominent in the 7--14 keV range and broadening of off-pulse appearing at higher energies. The energy-average pulse profile is overplotted as a dashed line.}
\label{fig:dyn-pp}
\end{figure}

\begin{figure}
\centering
 \includegraphics[width=\linewidth]{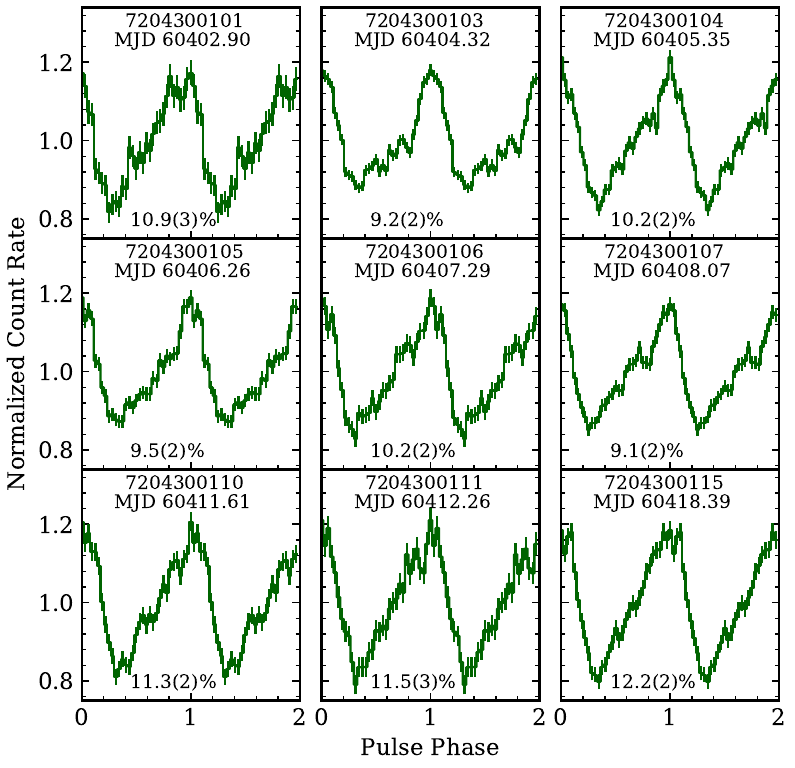}
\caption{Pulse profiles in the 0.5--10 keV energy range from individual \nicer\ observations, folded using the respective spin frequencies reported in Table \ref{tab:nicerspin}. The profiles are phase-aligned such that the main peak is centered around phase 1. The corresponding obsID, date (in MJD), and pulsed fraction ($PF_{\rm rms}$) are mentioned at the top and bottom of each panel, respectively. }
\label{fig:ni-profile}
\end{figure}


\section{Results}
\label{res}

\subsection{Timing analysis}
\label{timing}

Fig. \ref{fig:lc} shows the \lxp\ light curves of \src\ in the 3--25 keV energy range, with the bottom panel displaying the hardness ratio, defined as the ratio of count rates in the 8--25 keV and 3--8 keV energy ranges.  
The light curve exhibits pronounced flaring activity within the first 40 ks of the observation. To systematically identify these flares, we analyzed each good time interval (GTI) using the \texttt{find\_peaks} routine from the \textsc{scipy} package \citep{Virtanen20}, applying a detection threshold of 75 counts s$^{-1}$. We identified a total of seven flares, with durations ranging from 400 to 700 s, as highlighted by shaded regions in Fig. \ref{fig:lc}. During these events, the peak count rate increased by a factor of $\sim$1.5--2 relative to the persistent level. The bottom subplot highlights a representative segment of the light curve featuring one such flare, where the count rate increased by a factor of $\sim$2 during the peak of the flare compared to the pre- and post-flare count rates. Despite this variability in intensity, the hardness ratio remains largely constant throughout the \astrosat\ observation, suggesting little to no associated spectral variation during the flares.

We also examined the \nicer\ light curves for similar flaring activity. While the short duration of individual snapshots (400--600 s) made systematic identification challenging, we found evidence of significant count rate enhancements in a few cases. For instance, in observation ID 7204300102 and two snapshots of 7204300103, the 0.5--10 keV count rate increased from $\sim$30--40 counts \psec\ to over 90 counts \psec. Although the decay phase of these events was not captured due to GTI limitations, visual inspection indicates that the flares likely lasted more than 400 s.

To investigate the aperiodic and periodic timing behaviour, we extracted a power density spectrum (PDS) from the 3--25 keV LAXPC light curve, binned at 0.5 s. The PDS was averaged over 17 segments of 4096 s each and then logarithmically rebinned with a factor of 1.02. The resulting PDS was Poisson noise subtracted and rms normalized, is shown in Fig. \ref{fig:pds}. The PDS shows prominent peaks at the fundamental spin frequency of the neutron star, $\sim$0.124 Hz (corresponding to a spin period of 8.03 s) and its harmonics superimposed on a continuum of red noise. The noise continuum can be best described by a combination of three Lorentzians \citep{Belloni02, Reig08, Sharma24}, representing peaked noise around 1, 10, and 500 mHz, likely reflecting variability in the accretion environment on multiple timescales. Stochastic variability on timescales of around hundred seconds is also evident in the light curve itself (bottom panel of Fig. \ref{fig:lc}). The three Lorentzian components had rms of 11.8$\pm$1.2\%, 12$\pm$1\%, and 20.5$\pm$0.5\%, respectively.

To determine the spin frequency of the neutron star, we first applied the epoch-folding technique using the \texttt{efsearch} task \citep{Leahy87} on the barycentre-corrected photon arrival times from the \nicer\ and \astrosat\ observations. For each \nicer\ observation, the best spin frequency was determined, and uncertainties were estimated using the bootstrap method \citep{Boldin2013}, by simulating 1000 light curves following the method of \citet{Sharma23-lmc}. For observations with at least three snapshots, we also confirmed the spin frequencies using a phase-connection approach. Coherent pulsations at $\sim$8 s were detected across all \nicer\ observations, although some lower-exposure observations did not permit precise spin tracking. The resulting spin frequencies are listed in Table \ref{tab:nicerspin}. 

For the \astrosat\ observation, an initial spin estimate was similarly refined using a phase-connection technique, which revealed a gradual phase drift in the pulse profile, indicative of a non-zero spin frequency derivative. This barycentric analysis yielded a spin frequency of $\nu_0 = 0.12454877 (15)$ Hz and a negative spin frequency derivative, suggesting an apparent spin-down. However, this result does not account for orbital motion and is therefore not physically meaningful. 

The top panel of Fig. \ref{fig:spin} shows the spin frequency evolution of \src, combining measurements from \nicer, \astrosat, \nustar\ \citep[see,][]{Yang25} and \emph{Fermi}-GBM \citep{Malacaria20GBM}. The spin frequency exhibits a clear sinusoidal modulation, indicative of Doppler shifts due to orbital motion. To account for this, we used a composite expression that accounts for both the intrinsic spin evolution and orbital Doppler shifts:
\begin{equation}
    \nu (t) = \nu_{\rm int} (t) - \nu_{\rm orb} (t)
\end{equation}
where, $\nu_{\rm orb} (t)$ represent the frequency shift caused by the Doppler effect \citep{Galloway05}, and $\nu_{\rm int} (t)$ represents the intrinsic frequency of the source given by
\begin{equation}
    \nu_{\rm int} (t) = \nu_0 + \dot{\nu} (t-T_0) + \frac{1}{2} \ddot{\nu} (t-T_0)^2,
\end{equation}
where $\nu_0$, $\dot{\nu}$ and $\ddot{\nu}$ are the spin frequency, spin derivative and second derivative, respectively, at the reference time $T_0$. 

We modeled $\nu_{\rm orb} (t)$, using the orbital parameters from \citet{Yang25}. Since there exists a degeneracy in the orbital solutions proposed by \citet{Yang25}, primarily due to uncertainty in the exact number of orbital cycles that occurred between the two successive outbursts. We adopted their Solution-II, with an orbital period of 24.38 days, as it is closer to the reported optical period of 24.43 days \citep{Vasilopoulos14b}. We did not attempt to fit the orbital parameters, as our data do not provide tighter constraints on the orbit than those already reported.

We then applied orbital correction using Solution II to the \nicer, \astrosat, and \nustar{} data and repeated the timing analysis. For the \astrosat{} dataset, this yielded an updated spin frequency of $\nu_0 = 0.12456285(15)$ Hz and a spin derivative of $\dot{\nu} = 2.4 (2) \times 10^{-11}$ Hz s$^{-1}$, consistent with accretion-driven spin-up. The orbital-corrected spin frequencies are listed in Table \ref{tab:nicerspin}. The middle panel of Fig. \ref{fig:spin} shows the orbital-corrected spin frequencies from \nicer, \astrosat, and \nustar. Orbital-corrected spin frequencies from \emph{Fermi}-GBM were excluded, as they were derived using a different orbital solution than the one adopted in this work. A clear increasing trend in spin frequencies is observed, supporting spin-up during the outburst \citep{Yang25}. 

The spin trend was modeled using both linear and quadratic forms of the intrinsic spin evolution ($\nu_{\rm int} (t)$). A linear fit ($\ddot{\nu}=0$) to all points yielded a poor fit statistic ($\chi^2$/dof=235/9), indicating that a constant spin-up rate cannot adequately describe the observed evolution. However, if the final \nicer\ point at MJD 60418 is excluded, the linear model (dashed line) provides a reasonable fit with $\chi^2$/dof=13.5/9. In contrast, a quadratic model (solid line) better captures the full evolution ($\chi^2$/dof=5/8). The quadratic term revealed a measurable second derivative of the spin frequency, $\ddot{\nu} = -2.11(14)\times10^{-17}$ Hz s$^{-2}$, indicating a changing spin-up rate during the outburst. The bottom panel of Fig. \ref{fig:spin} shows residuals of the quadratic model with all observations and the linear model excluding the final point, clearly illustrating the improved fit provided by the second-order term. For comparison, the top panel also overplots the composite model $\nu (t)$ using the best-fit intrinsic spin parameters and the fixed orbital solution. The consistency between the composite model and the observed spin frequency evolution validates the orbital correction and confirms the presence of an evolving spin-up torque during the outburst. Table \ref{tab:spin} summarizes the spin parameters inferred from the orbital-corrected timing analysis. The spin parameters inferred from the broadband \astrosat{} observation are in agreement with the long-term trend seen across the outburst. 

We also tested Solution-I from \citet{Yang25}, with an orbital period of 23.92 days. The spin frequencies obtained from individual observations and the overall spin evolution were found to be broadly consistent within uncertainties, with those derived using Solution-II.


\begin{figure*}
\centering
  \includegraphics[width=\linewidth]{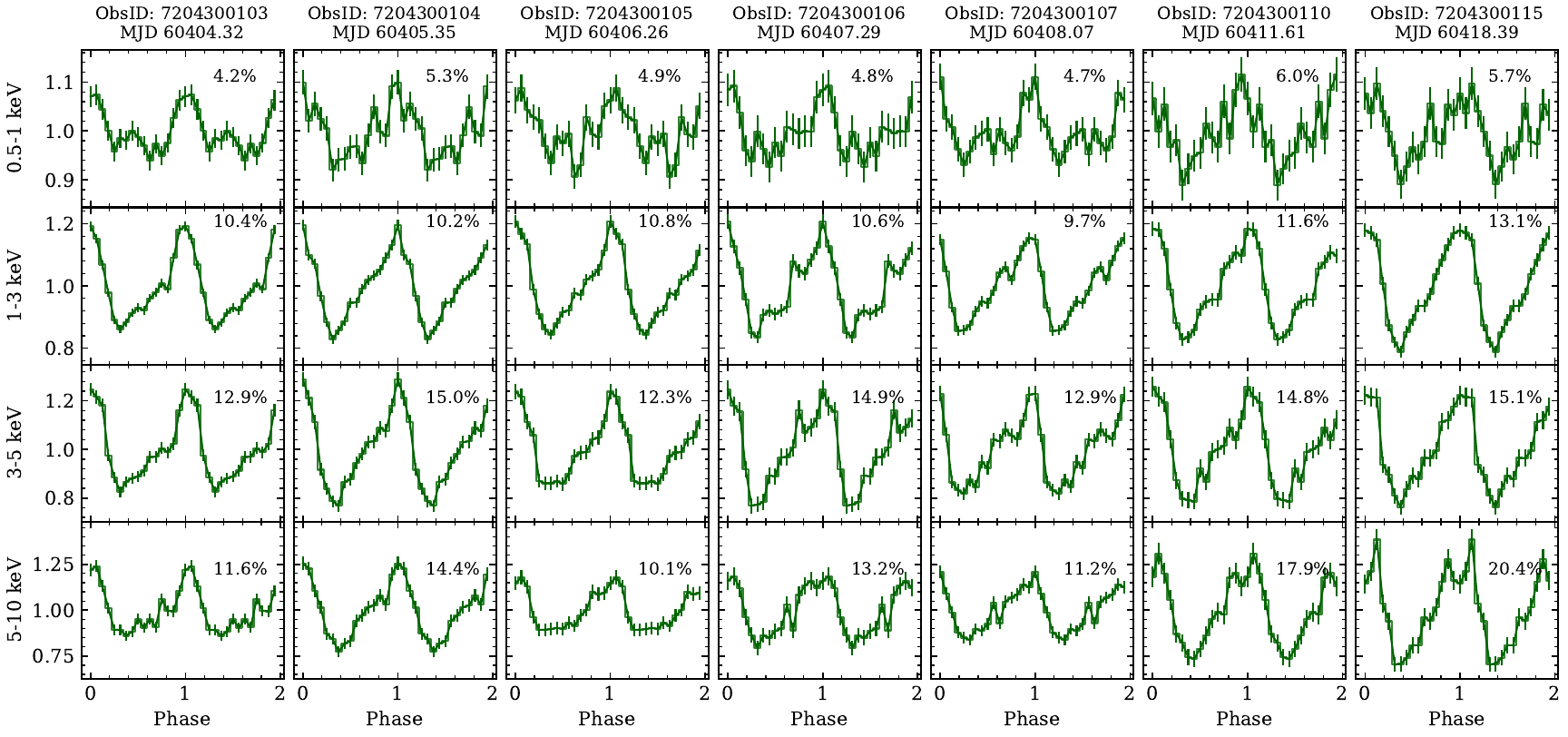}
 \caption{The energy-resolved pulse profile from \nicer\ observations. The corresponding obsID and date (in MJD) are mentioned on top of each figure. The corresponding pulsed fraction ($PF_{\rm rms}$) in each selected energy range is mentioned in each subpanel.}
\label{fig:ni103}
\end{figure*}

\begin{figure}
\centering
  \includegraphics[width=0.9\linewidth]{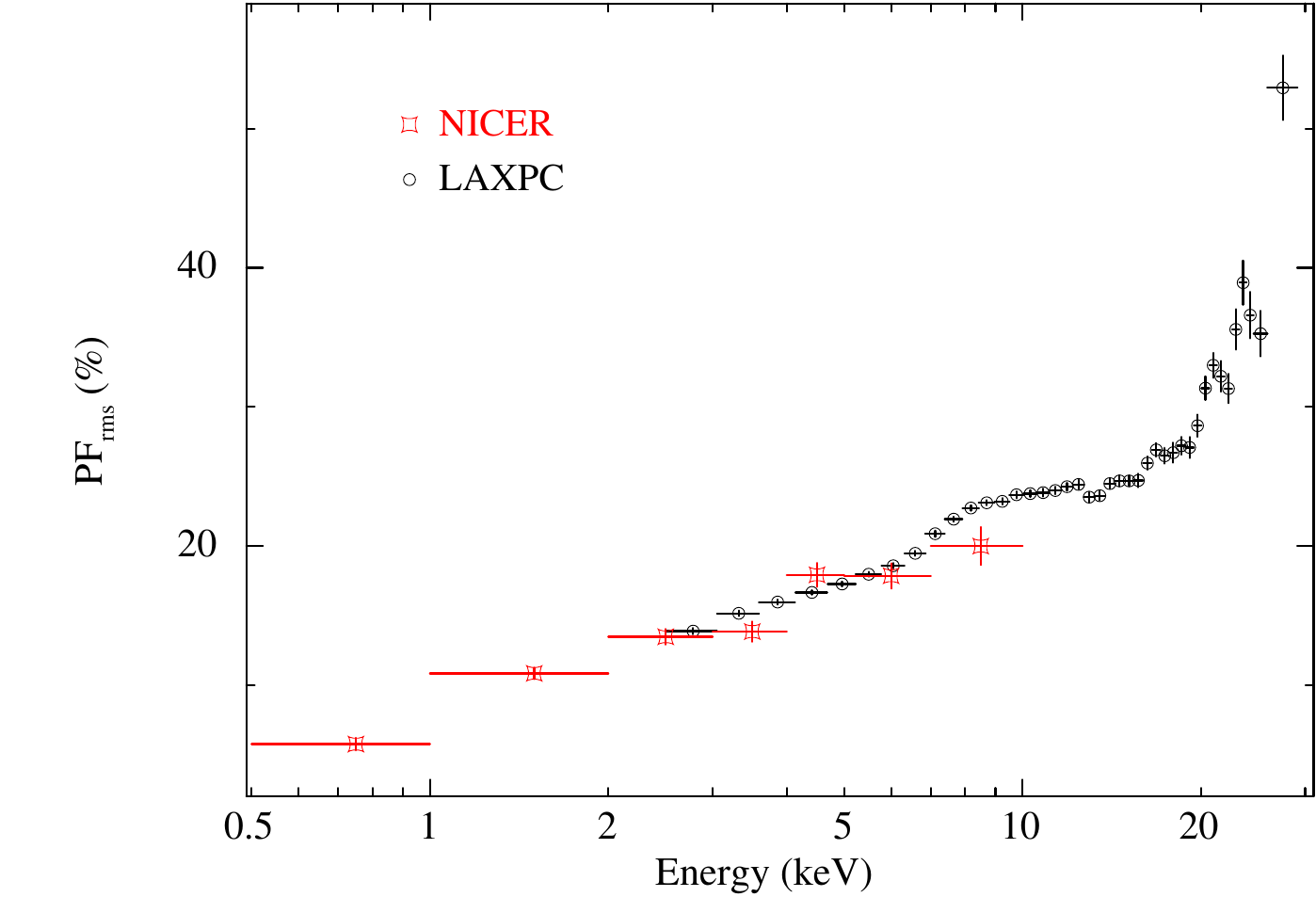}
 \caption{The rms pulsed fraction as a function of energy, estimated using simultaneous \astrosat/LAXPC and \nicer{} observations. The pulsed fraction shows a clear increasing trend with energy.}
\label{fig:pf}
\end{figure}

\begin{figure*}
\centering
\includegraphics[width=0.49\linewidth]{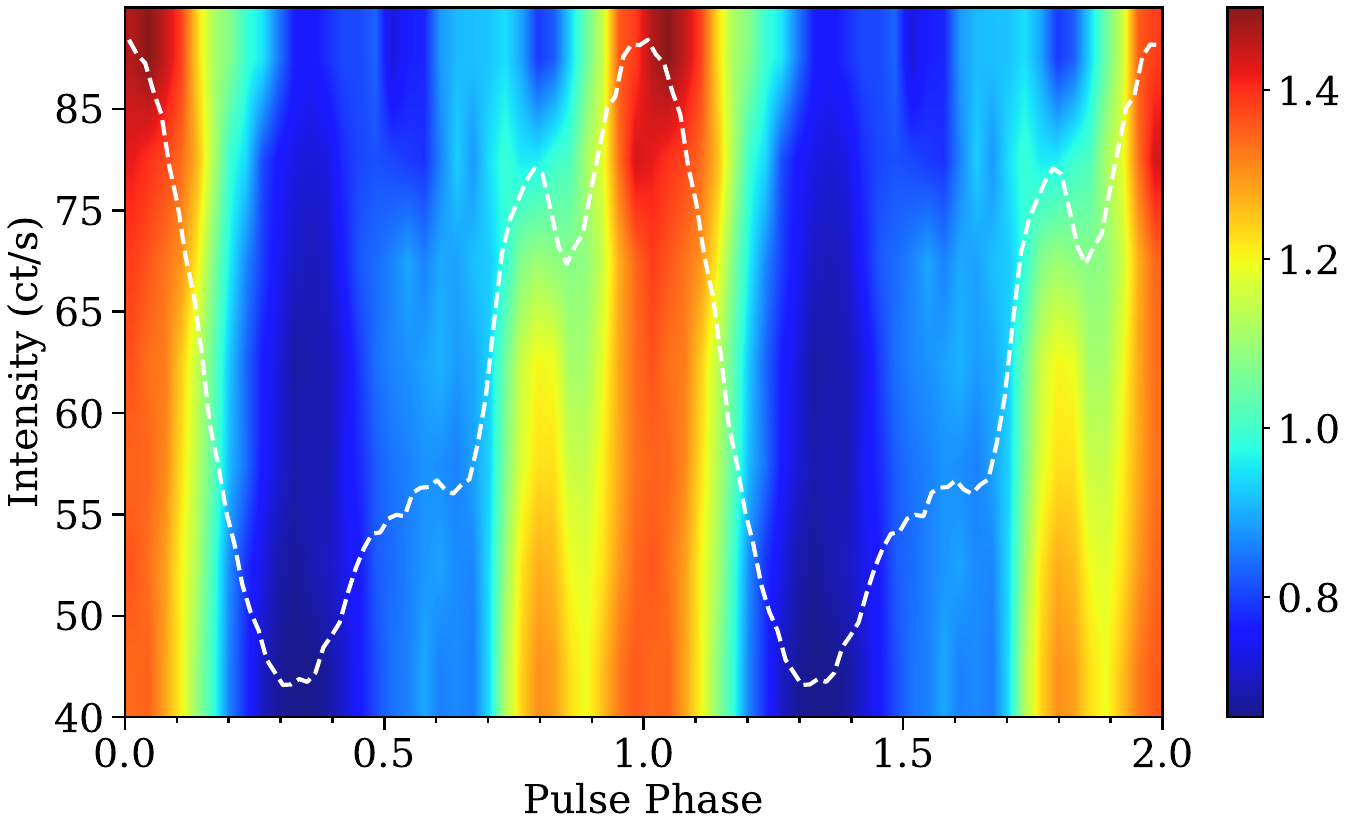}
 \includegraphics[width=0.49\linewidth, height=5.5cm]{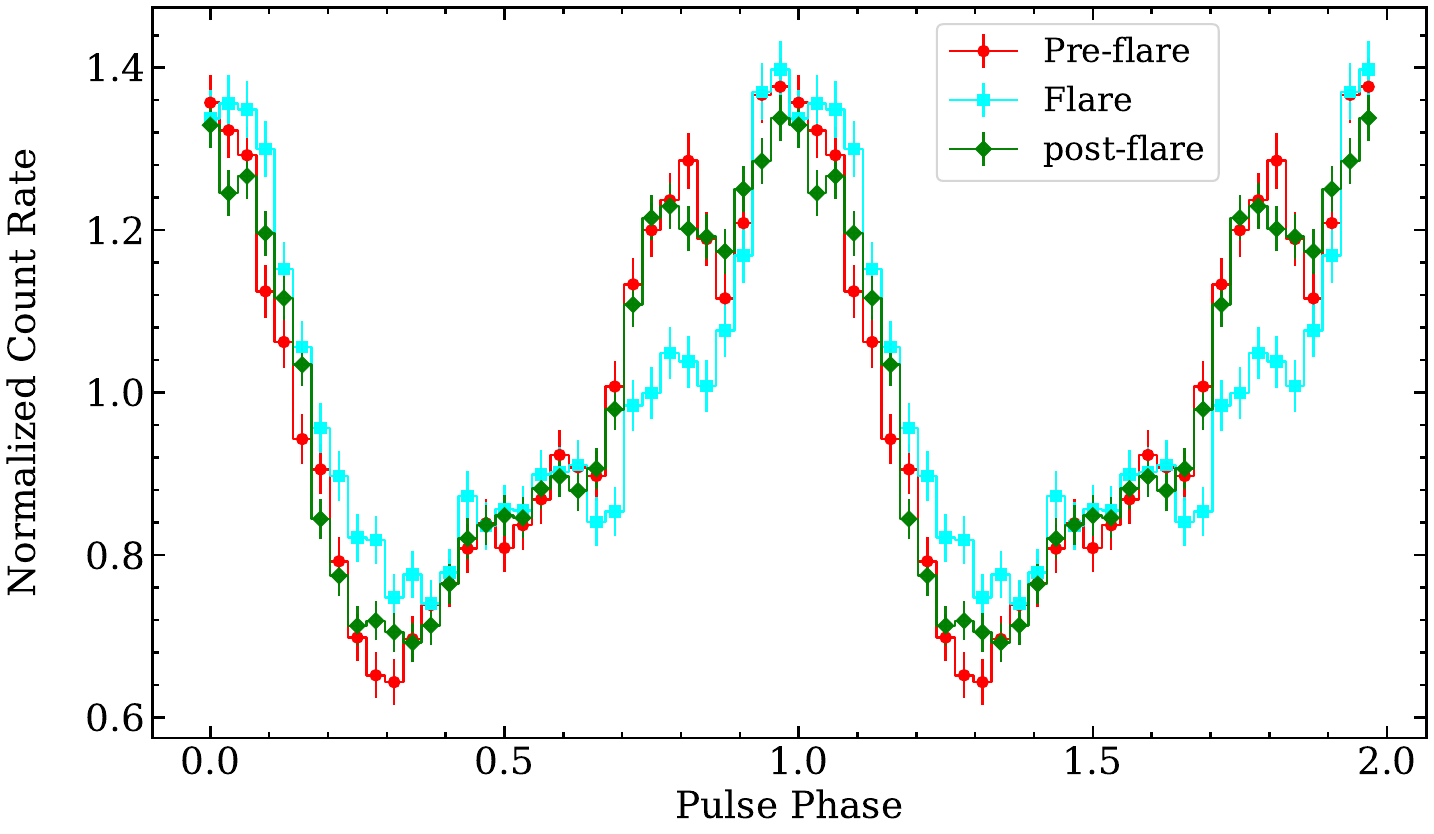}
 \caption{\textit{Left:} Intensity-resolved dynamic pulse profiles from \astrosat/LAXPC data during the 2024 outburst. The colormap shows normalized pulse shapes in different intensity intervals, revealing notable structural changes in pulse morphology. Intensity-average pulse profile is overplotted as a dashed line for comparison.
\textit{Right:} Pulse profiles extracted for three distinct flux states identified in Fig.~\ref{fig:lc}: pre-flare, flare, and post-flare intervals. The secondary peak is clearly visible before and after the flare but disappears during the flare, suggesting a change in the emission geometry.}
\label{fig:dyn-flux-pp}
\end{figure*}


\subsubsection{Pulse profiles}

We created the pulse profile of \src\ using orbital-corrected \lxp\ data folded with the respective spin frequency and derivative listed in Table \ref{tab:spin}. The energy-average pulse profile is shown in Fig. \ref{fig:dyn-pp} with a dashed line. The profile is clearly asymmetric, exhibiting a secondary minor peak $\sim$0.2 in phase before the main peak, and a left-side wing preceding the minor peak. 
To examine the energy dependence of the pulse morphology, we extracted background-corrected light curves in multiple energy bands from the \lxp\ observation. Fig. \ref{fig:dyn-pp} shows a colormap illustrating the evolution of pulse profile shape across energy. No significant pulsed signal was detected above 29 keV, likely due to background dominance and limited photon statistics. The secondary peak is most pronounced in the intermediate energy range ($\sim$7--14 keV). At low energies (3--4 keV), the source showed a single-peaked asymmetric pulse profile. At higher energies (17--20 keV), the wing emission before the minor peak diminishes, giving way to a broadening of the off-pulse region, and a noticeable excess appears near phase 0.4--0.5, creating twin dips on either side of this excess. The profile above 23 keV shows a deeper dip around phase 0.6, indicating significant evolution of the emission geometry with energy. For clarity, we also present the pulse profiles in the selected energy range of 3--4, 4--7, 7--14, 14--20, 20--23, and 23--29 keV in Fig. \ref{fig:e-res-pp}.

Fig. \ref{fig:ni-profile} shows the pulse profiles from individual, orbital-corrected \nicer\ observations in the 0.5--10 keV energy range, corresponding to epochs where spin frequency measurements were performed. Each profile was folded using the respective spin frequency listed in Table~\ref{tab:nicerspin}. The resulting profiles are predominantly single-peaked and asymmetric, with minor substructures that evolve with time. Next, we generated energy-resolved profiles from the \nicer\ observations with exposures exceeding 1~ks. Fig. \ref{fig:ni103} shows the resulting profiles in the energy range of 0.5--1, 1--3, 3--5, and 5--10 keV. These profiles also showed clear energy dependence with evolving minor structures. 

To quantify the pulse modulation, we calculated the root-mean-square (rms) pulsed fraction, $PF_{\rm rms}$, following the definition from \citet{Wilson18}:
\begin{equation}
PF_{\rm rms} = \frac{1}{\bar{p} \sqrt{N}} \left[\sum \limits_{i=1}^N (p_{i} -\bar{p})^2\right]^\frac{1}{2},
\end{equation}
where $N$ is the number of phase bins, $\Bar{p}$ is the mean count rate, and $p_i$ is the count rate in the $i$-th phase bin. The $PF_{\rm rms}$ values computed from individual pulse profiles from \nicer are mentioned in each subpanel of Fig. \ref{fig:ni-profile} and \ref{fig:ni103}, with the corresponding observation dates noted at the top. The $PF_{\rm rms}$ varied between $\sim$9--12\% in the 0.5--10 keV energy range. The energy-resolved profiles showed large dynamic variation of pulsed fraction, with increaing trend with energy. The $PF_{\rm rms}$ ranged from $\sim$4--6\% in the 0.5--1 keV, $\sim$10--13\% in 1--3 keV, $\sim$12--15\% in 3--5 keV, and $\sim$10--20\% in 5--10 keV.

For the \lxp\ data, the $PF_{rms}$ was found to be $\sim$22\% in the 3--25 keV energy range. To further examine the energy dependence of pulse modulation, we computed $PF_{\rm rms}$ across multiple energy bands using simultaneous \astrosat\ and \nicer\ observations. The resulting $PF_{\rm rms}$ as a function of energy is shown in Fig. \ref{fig:pf}, revealing a clear increasing trend with energy, consistent with findings from other studies \citep{Lutovinov09, Ferrigno23, Sharma24, Yang25}.

Flaring episodes observed during the \astrosat{} observation prompted an investigation into the intensity dependence of the pulse profiles. We selected one of the segments with flare and extracted light curves for the pre-flare, flare, and post-flare intervals as identified in the inset panel of Fig. \ref{fig:lc}. The corresponding pulse profiles are shown in the right panel of Fig. \ref{fig:dyn-flux-pp}. At lower or persistent intensities (pre- and post-flare), the secondary peak near phase $\sim$0.8 is clearly visible, but it vanishes entirely during the flaring phase. Next, we generated different intensity-filtered light curves using data from the entire observation, and the left panel displays the dynamic intensity-resolved colormap of the pulsed emission, highlighting that the secondary peak does not merge with the first peak but rather disappears during the flares. At low intensities ($\lesssim$65 counts s$^{-1}$), the secondary peak is well defined, but it diminishes at higher intensities and is absent during bright flaring phases. This behavior suggests changes in the emission geometry during flaring intervals, potentially driven by variable accretion from clumpy or inhomogeneous flows. The $PF_{\rm rms}$ also varies with intensity: decreasing from 24.4\% to 20.7\% between the low and intermediate count-rate regimes (40 to 65 counts s$^{-1}$), and then rising to 23.8\% for $>$85 counts s$^{-1}$, indicating non-monotonic modulation likely tied to accretion dynamics.



\subsection{Spectral analysis}
\label{spectral}

We performed spectral fitting using \textsc{xspec} version 12.14.1 \citep{Arnaud}. A distance of 50 kpc was assumed for the source, consistent with its location in the LMC \citep{Pietrzynski13}. Given the low metallicity environment of the LMC, the photo-electric absorption was modelled as a combination of Galactic foreground absorption and an additional column density accounting for both the interstellar medium of the LMC and the absorption local to the source. The Galactic absorption was fixed at $N_{\rm H}^{\rm GAL} = 6.44 \times 10^{20}$ cm$^{-2}$ \citep{DLmap90}, with elemental abundances set according to \citet{Wilms}. The intrinsic LMC absorption, $N_{\rm H}^{\rm LMC}$ was treated as a free parameter in the fit, with sub-solar abundances (0.49 for elements heavier than helium) adopted based on \citet{Rolleston02}, following \cite{Vasilopoulos14b, Vasilopoulos16}. Unless otherwise specified, all uncertainties and upper limits on the spectral parameters are quoted at the 90\% confidence level for one parameter of interest.

\subsubsection{NICER+AstroSat}

The \nicer\ observation (OBS ID: 7204300110) is contemporaneous with \astrosat. To study the broadband spectrum of \src{}, we performed a combined spectral analysis of \nicer, SXT, and LAXPC20 data. We consider the LAXPC20 data up to 50 keV for the spectral fitting. The SXT data in the energy range of 1--7 keV are used for the combined spectral fitting. We added a constant component representing the cross-calibration between the \nicer, SXT, and LAXPC20 instruments. We also applied a gain correction for the SXT. The gain slope was fixed to 1.0, and the offset was allowed to vary. The gain offset was found to be $-$32 eV. A systematic uncertainty of 2\% was used during spectral fitting. 

The hard X-ray continuum spectrum of X-ray pulsars can be fitted by a phenomenological powerlaw-like shape with an exponential high-energy cut-off that originates from the accretion column \citep[e.g.,][]{Coburn02, Maitra18}. We modelled the combined broadband spectra of \src\ in the energy range of 0.5--50 keV using an absorbed powerlaw with a high-energy cutoff model. This model alone provided a statistically unacceptable fit ($\chi^2$/dof = 545/231), with significant residuals seen below 1 keV (soft excess) and around 30 keV (absorption-like dip). Soft excesses are commonly observed in the spectra of X-ray pulsars and may arise from several physical mechanisms, including reprocessed emission from the inner accretion disc, hot plasma near the magnetosphere, or the neutron star surface, and may also be shaped by partial covering absorption \citep{Hickox2004}. Additionally, a CRSF at $\sim$32 keV has been previously reported in this source \citep{Tendulkar14, Yang25}. We therefore extended the model to include a thermal blackbody component (\texttt{bbodyrad}) for the soft excess and a multiplicative CRSF component \citep[\texttt{cyclabs};][]{Makishima90, Mihara90} to model the absorption dip. This improved the fit significantly ($\chi^2$/dof = 286/226). 

\begin{figure}
\centering
 \includegraphics[width=0.99\linewidth]{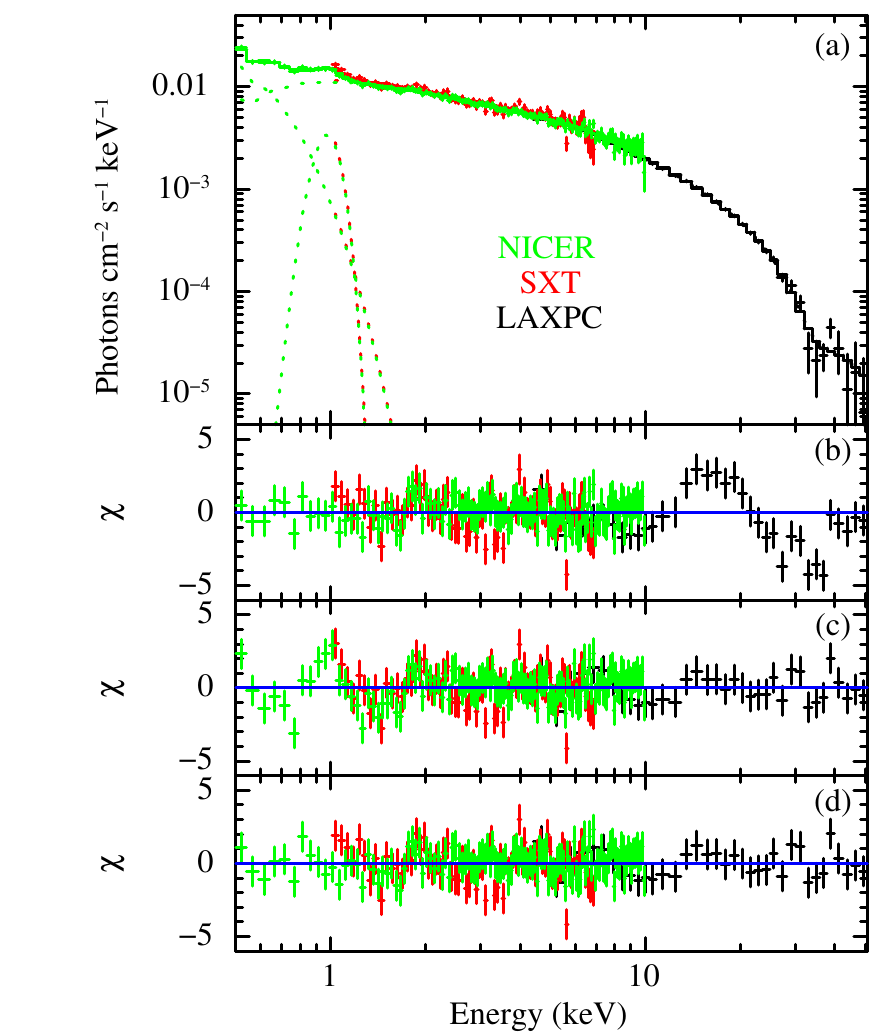}
 \caption{(a): Unfolded broadband X-ray spectrum of \src\ from \nicer, SXT and LAXPC, fitted with the model \texttt{tbabs*tbvarabs*(bbodyrad + Gaussian + highecut*powerlaw*cyclabs)}. (b) Residuals ($\chi$ = (data-model)/error) after fitting without the \texttt{cyclabs} model. (c) Residuals after fitting without the 1 keV Gaussian emission line. (d) Residuals from the final best-fit model.}
\label{fig:spec}
\end{figure}

Nonetheless, residuals near $\sim$1 keV suggested the presence of an emission line (see Fig. \ref{fig:spec}c). Introducing a Gaussian emission line component at this energy further improved the fit ($\chi^2$/dof = 239/224). When left free, the line width was found to be relatively broad ($\sim$0.17 keV), which strongly influenced the soft excess parameters, particularly lowering the blackbody temperature to 0.07 keV and increasing its normalization. To mitigate this degeneracy and obtain physically meaningful parameters, we fixed the Gaussian line width at the \nicer\ spectral resolution of 85 eV. The observed broadening may be due to a blend of unresolved emission lines around $\sim$1 keV, such as those from Ne \textsc{ix/x} or Fe L-shell transitions. Fixing the line width avoids degeneracies with the soft excess continuum, yields a stable blackbody component with temperature $\sim$0.09 keV, and effectively models the residual structure without overestimating the soft flux. We also explored an alternative interpretation of the soft excess using a partial covering absorption model instead of the blackbody component. However, this resulted in a significantly poorer fit ($\chi^2$/dof = 274/224), disfavoring this scenario.  

We evaluated the statistical significance of the additional components: the blackbody, Gaussian line, and CRSF, using the \texttt{simftest} routine in \textsc{xspec}, based on 10,000 Monte Carlo simulations. In all three cases, the false alarm probability was found to be $<10^{-4}$, confirming their significance at greater than $3\sigma$ confidence. Furthermore, we found no strong evidence of an Fe K$\alpha$ line in the spectrum, and derived an upper limit of 121 eV on its equivalent width, assuming a fixed line energy of 6.5 keV and a line width of 0.27 keV \citep{Yang25}.

The best-fit spectral parameters are listed in Table~\ref{tab:fitstat}, and the best-fit spectrum along with model components and residuals is shown in Fig.~\ref{fig:spec}. We obtained an unabsorbed flux of $\sim 7\times 10^{-10}$ erg cm$^{-2}$ s$^{-1}$ in the 0.5--50 keV range. Extrapolating the model over the 0.1--100 keV band, the flux was estimated to be $7.5\times 10^{-10}$ erg cm$^{-2}$ s$^{-1}$, corresponding to an X-ray luminosity of $2.24\times 10^{38}$ erg s$^{-1}$ for a source distance of 50 kpc \citep{Pietrzynski13}. 

We also tested alternative continuum models such as \texttt{cutoffpl}, \texttt{fdcut*powerlaw}, or \texttt{NPEX}, but these provided poorer fits than the \texttt{powerlaw*highecut} model adopted here. 
Phase-resolved spectroscopy was not performed due to limited photon statistics. 


\renewcommand{\arraystretch}{1.2}
\begin{table}
	\centering
	\caption{Best-fit spectral parameters of \src\ using the model \texttt{tbabs*tbvarabs*(blackbody + Gaussian + highecut*powerlaw*cyclabs)}. All errors and upper limits reported in this table are at a 90\% confidence level ($\Delta\chi^2=2.7$).}
	\label{tab:fitstat}
	\resizebox{0.75\linewidth}{!}{
	\begin{tabular}{lcc} 
		\hline
	 Model & Parameters & Values  \\
		\hline
    \texttt{tbvarabs} & $N_{\rm H}^{\rm LMC}$ ($10^{21}$ cm$^{-2}$) & $1.5^{+0.9}_{-1.0}$  \\
    
    \texttt{Bbodyrad} & $kT_{\rm BB}$ (keV) & $0.092_{-0.007}^{+0.010}$   \\  
                   & Norm ($10^{4}$) & $5.0_{-3.6}^{+8.0}$  \\
                   & $R_{\rm BB}$ (km) & $1114_{-400}^{+900}$  \\  
                   
    \texttt{highecut} & $E_{\rm cut}$ (keV) & $5.8 \pm 0.5$  \\ 
                   & $E_{\rm fold}$ (keV) & $11.8^{+0.9}_{-0.7}$  \\ 
                   
    \texttt{powerlaw} & $\Gamma$ & $0.70 \pm 0.04$ \\
                   & Norm ($10^{-2}$) & $1.46^{+0.08}_{-0.07}$  \\   
                   
    \texttt{Gaussian} & $E$ (keV) & $0.97 \pm 0.02$   \\ 
                   & Norm ($10^{-3}$) & $1.0 \pm 0.3$  \\ 
                   & EQW (eV) & $67 \pm 23$\\
                   
    \texttt{Cyclabs} & $E$ (keV) & $33.0^{+2.9}_{-1.4}$   \\ 
                   & width (keV) & $7.7^{+6.1}_{-3.1}$  \\  
                   & Depth & $1.1^{+0.4}_{-0.2}$  \\ 
                   
     \texttt{Cons} & $C_{\rm SXT}$ & $1.05 \pm 0.02$ \\ 
                   & $C_{\rm LAXPC}$ & $0.98 \pm 0.02$ \\ 
                   & $C_{NICER}$ & 1 (fixed) \\
                   
                   & Flux$_{\rm 0.5-50~keV}^a$ & $6.98 ~(14) \times 10^{-10}$  \\ 
                   & Flux$_{\rm 0.1-100~keV}^a$ & $7.50 ~(17) \times 10^{-10}$  \\   

                   & L$_{\rm 0.5-50~keV}^b$ & $2.09 ~(4) \times 10^{38}$  \\ 
                   & L$_{\rm 0.1-100~keV}^b$ & $2.24 ~(5) \times 10^{38}$  \\                     
\hline                   
                   & $\chi^2$/dof & 239/224  \\                    
		\hline
	\end{tabular}}
    \tablefoot{\tablefoottext{a}{Unabsorbed flux in the units of \erg.} 
    \tablefoottext{b}{X-ray Luminosity in the units of \lum.}}
\end{table}

\subsubsection{Individual NICER spectrum}

The continuum of individual \nicer\ spectra obtained during the outburst is well described by a model consisting of an absorbed powerlaw, a soft thermal blackbody component, and a narrow Gaussian emission line at $\sim$1 keV. 
We also attempted to fit the spectra with a powerlaw with a cut-off, but no cut-off is required within the \nicer\ energy band, except for 7204300115, where a cut-off was observed around 7 keV. 

\begin{figure}
\centering
 \includegraphics[width=0.85\linewidth, height=15cm]{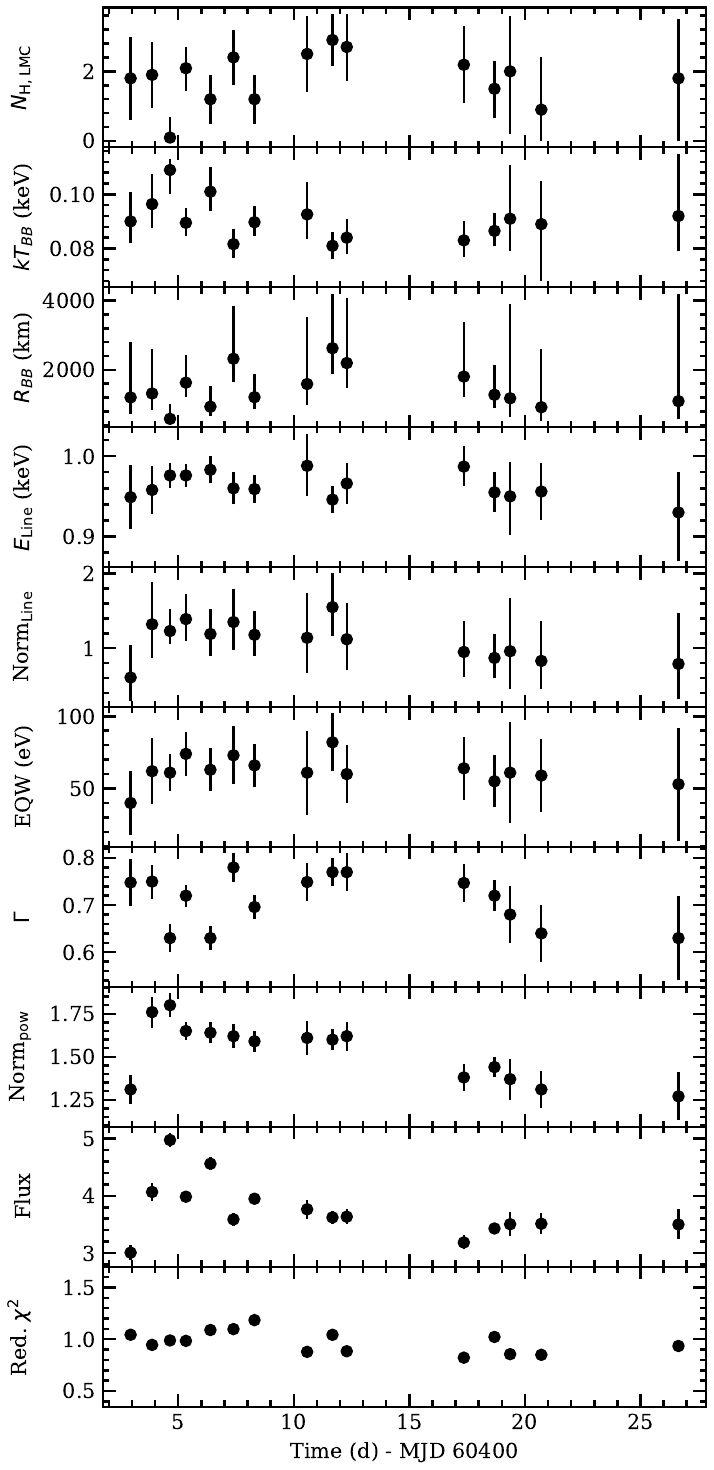}
 \caption{Temporal variation of best-fit spectral parameters of \src\ from individual \nicer\ observations during the 2024 outburst. The panels show (from top to bottom): LMC Column density ($N_{\rm H}^{\rm LMC}$) in units of $10^{21}$ \pcm, blackbody temperature ($kT_{\rm BB}$), blackbody radius ($R_{\rm BB}$) assuming a distance of 50 kpc, emission line centroid energy ($E_{\rm line}$), normalization of emission line ($\times10^{-3}$), equivalent width of emission line, photon index ($\Gamma$), normalization of powerlaw ($\times10^{-2}$), unabsorbed flux in the 0.5--10 keV range in units of $10^{-10}$ \erg\ and reduced $\chi^2$ of best-fit, repsectively.}
\label{fig:spec-nicer2}
\end{figure}

Fig. \ref{fig:spec-nicer2} presents the temporal evolution of the best-fit spectral parameters across all \nicer\ observations. The LMC column density ($N_{\rm H}^{\rm LMC}$) remained stable, with a mean value of $\sim 1.8 \times 10^{21}$ cm$^{-2}$. The photon index exhibited modest stochastic variations, with a mean value of $\Gamma \sim 0.71$ and a range of 0.63 to 0.78, consistent with values reported by \swiftxrt, \emph{EP}, and \emph{LEIA} observations \citep{Yang25}. After MJD 60417, the photon index showed a decreasing trend, indicating spectral hardening. However, comparable low values were also present during the initial phase of the outburst, likely reflecting stochastic fluctuations rather than a sustained trend. The normalization of the power-law component steadily decreased over time. 
Other spectral parameters did not display significant systematic trends. The unabsorbed flux varied by approximately $\pm$25\% over the monitoring period. The soft thermal excess was persistently detected at a temperature of 0.08--0.11 keV, with an inferred emission radius of the order of $10^3$ km, consistent with emission from the inner accretion disc. Additionally, a broad emission line at $\sim$1 keV was detected in most \nicer\ spectra, with equivalent widths ranging from $\sim$40 to 80 eV.


\section{Discussion}
\label{dis}

In this work, we presented a comprehensive temporal and spectral study of the Be/X-ray binary pulsar \src{} using data from \astrosat{} and contemporaneous monitoring with \nicer{} during its 2024 outburst. We also used \nicer{} observations to track the evolution of the neutron star’s spin and spectral properties over the course of the outburst.

\subsection{Pulse properties and energy dependence}

We detected strong pulsations at a spin frequency of $\sim$124 mHz in both \astrosat\ and individual \nicer\ observations. The pulse profiles exhibited strong dependence on both energy and source intensity. The average profile from \lxp\ also showed an asymmetric, single-peaked structure, with a distinct minor peak around phase 0.2 preceding the main peak.

The energy-resolved pulse shapes provide insights into the geometry of the X-ray emitting regions and the evolution of the radiation beam pattern with energy and accretion dynamics. The energy-resolved profiles from \lxp\ showed strong energy dependence. At low energies (3--4 keV), the profile is asymmetric and lacks a secondary peak. In the intermediate band ($\sim$7--14 keV), the profile became more complex, exhibiting a prominent secondary peak. At higher energies, the profile became single-peaked with broader off-pulse emission and a minor excess around phase 0.5 in the 17--20 keV band. Above 23 keV, the profile exhibited a deeper dip near phase 0.6 and lacked the excess at phase 0.5. The overall pulse morphology and its energy dependence during the \astrosat\ observation are consistent with recent \nustar\ findings \citep{Yang25}. In particular, \nustar\ observations revealed a phase drift of the main peak by $\sim$0.15 from lower energies up to $\sim$30 keV, along with a gradual filling of the main trough at higher energies. A similar phase drift of $\sim$0.06 was observed in the \lxp\ data from low energies up to $\sim$25 keV, and a deeper dip appeared above 23 keV.

The profile from individual \nicer\ observations was asymmetric, single-peaked, with minor structures that evolved with both time and energy. The energy-resolved \nicer\ profiles exhibited notable variability in the 0.5--1 keV band, where reprocessed emission and contributions from Ne \textsc{ix}/Ne \textsc{x} and Fe L lines are expected to dominate. This variability could arise from azimuthally asymmetric reprocessing of the pulsed emission or from emission line blends produced in the irradiated inner disc. However, given the moderate spectral resolution of the data, it is not possible to unambiguously distinguish between a reprocessed continuum component and blended line emission.

Such energy-dependent evolution of pulse profiles is commonly observed in X-ray pulsars. Typically, profiles are more complex at lower energies and become simpler, single or double-peaked at higher energies, as seen in sources like Vela X--1, 4U 1626--67, and LMC X--4 \citep{Maitra13, Beri2014, Alonso-Hernandez22, Sharma2023-1626, Sharma23-lmc, Sharma25-1626}. At the lowest energies, where thermal reprocessed emission dominates (around $\sim$1 keV), the profiles can revert to a simpler single-peaked form, for example, in LMC X--4 \citep{Beri2017}. In contrast, \src\ exhibits a single asymmetric profile at low energies (above the thermal component), but shows complex multiple peaks at intermediate and higher energies, indicating a non-trivial beam pattern and possible contributions from different emission regions.

\subsection{Flares and pulse profile change}

Another prominent feature in the light curves of \src\ is the presence of pronounced aperiodic variability and flares, with a time-scale of hundreds of seconds. Flares are on average brighter than the persistent emission by a factor of $\sim$2. Similar flaring behavior has been observed in other accreting pulsars, including Vela X--1 \citep{Maitra13}, GRO J1744--28 \citep{Cannizzo97, Woods2000}, SMC X--1 \citep{Moon03}, Swift J1626.6--5156 \citep{Reig08-flare}, LMC X--4 \citep{Moon01, Beri2017, Shtykovsky2018}, A0538--66 \citep{Ducci19}, EXO 2030+375 \citep{Apparao91, Klochkov11}, 4U 1907+09 \citep{Doroshenko12}, and 4U 1901+03 \citep{James2011, Ji20}. These flares are generally attributed to instabilities in the accretion flow, such as Rayleigh–Taylor or magnetospheric instabilities, or to inhomogeneous (clumpy) stellar winds from the high-mass companion \citep{Taam88, Apparao91, Cannizzo97, Woods2000, Postnov08}. The brief flares observed in our data are consistent with the predictions of \citet{Romanova08} and \citet{Kulkarni08}, in which Rayleigh–Taylor interchange instabilities at the disk-magnetosphere boundary lead to the formation of penetrating tongues. These unstable streams produce transient, stochastic hotspots on the neutron star surface, producing the observed flaring and pulse variability. Despite strong flux variability during the flares, we found no significant change in the hardness ratio, suggesting that the spectral shape remains largely unaltered, similar to what has been observed in 4U 1901+03 \citep{James2011, Ji20}.

We also investigated the dependence of the pulse profile morphology on source intensity. At lower count rates (persistent or non-flaring, $\lesssim$65 counts s$^{-1}$), a secondary peak is clearly visible, but it disappears entirely during brighter flaring phases. This behavior can be interpreted as direct evidence for short-timescale changes in the beaming pattern, likely driven by variations in the accretion rate \citep{Ji20}. Similar to energy-resolved profiles, a phase drift of $\sim$0.07 was observed between the pulse profiles corresponding to intensities below 65 ct \psec\ and above 80 ct \psec. The pulsed fraction also varied non-monotonically with intensity, supporting the view that the pulsed emission is shaped by the luminosity-dependent beam pattern and accretion column geometry. Changes in the mass accretion rate alter the accretion structures and radiation transfer close to the polar cap, which reshape the radiation beam pattern and, consequently, the observed pulse profile is strongly influenced by luminosity \citep{Basko76, Becker12, Mushtukov15}. 

\subsection{Aperiodic variability}

X-ray pulsars demonstrate stochastic variabilities in their X-ray flux on different timescales, which manifest as red noise, quasi-periodic variation, and coherent pulsations in PDS \citep{Reig08}. It is believed that a significant fraction of the variability is formed in the accretion disc as a result of fluctuations of the local viscosity \cite[e.g.,][]{Lyubarskii97}. The PDS of \src{} also revealed red noise variability at multiple timescales, in addition to the prominent spin and harmonic peaks. The red noise was modeled using three Lorentzian components peaking at $\sim$1, 10, and 500 mHz, with fractional rms amplitudes of 11.8\%, 12\%, and 20.5\%, respectively. These components likely correspond to variations in the accretion flow at different spatial scales within the disc. The $\sim$500 mHz component may correspond to the Keplerian frequency at the inner edge of the accretion disc \citep{Revnivtsev09}, while the low-frequency variability (1--10 mHz) could be caused by stochastic accretion due to disc inhomogeneities or clumps causing flares.

\subsection{Spin evolution}

The spin evolution during giant outbursts offers valuable insight into accretion processes in BeXRBs. Accurate timing of pulsations across the orbit enables precise determination of orbital parameters through Doppler-induced modulations. As the neutron star accretes matter, the resulting torque can cause significant spin-up or spin-down, further modulated by orbital Doppler shifts. During the 2024 outburst, the barycentre-corrected spin frequencies of \src\ exhibited a nearly sinusoidal modulation due to orbital motion, superimposed with an intrinsic spin-up trend driven by accretion. After applying orbital corrections using the known binary parameters, the spin frequencies showed a steady increase over time, as consistently observed in both \astrosat, \nustar, and \nicer{} data. This behavior is analogous to the 2014 outburst, which also exhibited a sustained spin-up phase due to accretion \citep{Karaferias23, Yang25}.

Our long \astrosat\ observation, spanning over a day, enabled us to determine the intrinsic spin evolution using a phase-connection method. We measured a clear spin-up trend with a spin frequency derivative of $\dot{\nu} = 2.4(2) \times 10^{-11}$ Hz s$^{-1}$, consistent with spin-frequency evolution tracked across the outburst. Furthermore, we detected a second derivative of spin frequency, $\ddot{\nu} \sim -2 \times 10^{-17}$ Hz s$^{-2}$, indicating a gradual decline in the spin-up rate as the outburst progressed. This negative $\ddot{\nu}$ implies a weakening accretion torque over time, likely driven by a decrease in the mass accretion rate as the outburst decayed. A similar evolving spin-up rate has also been reported in other transient X-ray pulsars, for example, in 4U 1901+03 \citep{Galloway05}.

The \cite{Ghosh79-III} (GL) model is applicable to BeXB pulsars that accrete via a disc, especially in a strong accreting regime \citep{Bozzo09}. This model predicts spin-up rate ($\dot{\nu}$) in Hz s$^{-1}$
\begin{equation}
    \dot{\nu} = 1.37 \times 10^{-12}~ \mu_{30}^{2/7} n(\omega_s) \Big(\frac{R_{NS}}{10^6}\Big)^{6/7}  M_{1.4}^{-3/7} I_{45}^{-1} L_{37}^{6/7},
\end{equation}
where $\mu_{30}$ is magnetic moment in units of $10^{30}$ G cm$^3$, $I_{45}$ is moment inertia in $10^{45}$ g cm$^2$, $L_{37}=L_X/10^{37}$ \lum, $M_{1.4}=M_{NS}/1.4M_\odot$, $R_{NS}$ is the neutron star radius in cm, $L_{37}=L_X/10^{37}$ erg s$^{-1}$ is the X-ray luminosity, and $n(\omega_s)$ is the dimensionless accretion torque, which depends on the fastness parameter $\omega_s$, given by equations (10) and (16) of \citet{Ghosh79-III}. The magnetic moment is related to the surface dipole magnetic field $B$ at poles through $\mu = \tfrac{1}{2} B R_{NS}^3$. Assuming a canonical neutron star ($M_{NS} = 1.4 M_\odot$, $R_{NS} = 10^6$ cm), $B=3.6\times10^{12}$ G \citep[corrected for gravitational redshift;][]{Yang25} and a luminosity of $2.24 \times 10^{38}$ erg s$^{-1}$, we obtained $\dot{\nu} \simeq 2.6 \times 10^{-11}$ Hz \psec\ using the GL model, consistent with our measurement with timing analysis. Notably, the spin frequency derivative measured during the 2014 outburst was also in good agreement with the GL model \citep{Sugizaki17, Karaferias23}. 

\subsection{Spectral properties}

\src\ is observed to be accreting at an X-ray luminosity of $2.24\times10^{38}$ \lum, close to the Eddington limit. The broadband X-ray spectrum is well described by a powerlaw continuum with a high-energy cut-off dominating above $\sim$1 keV. At lower energies, we detect a soft excess emission component with a blackbody temperature of $\sim$0.09 keV. A similar soft excess at $\sim$0.1--0.3 keV has been observed in other X-ray pulsars such as SMC X-1 \citep{Paul2002}, LMC X-4 \citep{Paul2002, Naik2004, Sharma23-lmc}, Her X-1 \citep{DalFiume1998}, 4U 1626-67 \citep{Beri2015}, and in extra-galactic ultraluminous X-ray pulsars \citep{Kumar2025}. This thermal excess is likely to originate from reprocessed hard X-rays illuminating the inner accretion disc, which is truncated by the magnetosphere of the neutron star \citep{Paul2002, Hickox2004}. From the normalization of the blackbody component and assuming a spherical emission geometry, the radius of the emission region is estimated to be $1114_{-400}^{+900}$ km, for a distance of 50 kpc. 

The truncation radius of the disc is expected to coincide with the magnetospheric (or Alfvén) radius, where the magnetic pressure balances the ram pressure of accreting material \citep{Pringle72, Ghosh79-III}:
\begin{equation}
    R_M = k \Big( \frac{\mu^4}{2GM_{NS} \dot{M}^2} \Big)^{1/7}
\end{equation}
where $k$ is the dimensionless coupling constant between 0.5--1 \citep{Mushtukov22}, and $\dot{M}$ is the mass accretion rate inferred from the observed X-ray luminosity by $L_X = \eta \dot{M} c^2$ where $\eta \approx 0.2$ is the accretion efficiency. Assuming a canonical neutron star, we estimate $R_M \sim$1100--2100 km. This is in good agreement with the blackbody emission radius inferred from the spectral fit, further supporting a reprocessing origin for the soft excess.

In addition, we detect a broad emission feature at $\sim$0.97 keV with an equivalent width of $\sim$67 eV, likely originating from highly ionized Ne K or a blend of Fe L-shell transitions. Both the soft excess and this 1 keV line are consistently present in individual \nicer\ spectra. However, these features were not apparent in the \emph{LEIA} and \swiftxrt\ spectra, possibly due to limited statistics \citep{Yang25}. No statistically significant Fe K$\alpha$ emission was detected in our \astrosat+\nicer\ spectra, with an upper limit on the equivalent width of 121 eV. However, a subsequent \nustar\ observation conducted 1 day later revealed the presence of an Fe K$\alpha$ line with an equivalent width of $\sim$46 eV \citep{Yang25}. The upper limit derived from \astrosat+\nicer\ is not in conflict with the low equivalent width observed with \nustar. 

We also confirm the presence of a CRSF at $33.0^{+2.9}_{-1.4}$ keV, consistent with previous reports from \nustar\ \citep{Tendulkar14, Yang25}. The measured width and the optical depth of the CRSF are also found to be consistent with the \nustar\ measurements \citep{Yang25}. The photon indices obtained from the joint \astrosat\ and \nicer\ fit, as well as from individual \nicer\ spectra, are consistent with those reported by \citet{Yang25}.

\section{Conclusions}
\label{summary}

We have presented, for the first time, a detailed timing and broadband spectral investigation of the Be/X-ray binary pulsar \src\ during its 2024 outburst, based on coordinated observations with \nicer\ and \astrosat. Our results reveal several new aspects of the source’s accretion dynamics and emission behaviour: 

\begin{enumerate}
    \item Flaring activity: The source exhibited remarkable variability, including frequent short flaring episodes with durations of a few hundred seconds and flux enhancements by a factor of $\sim$2. Such flaring behaviour has not been reported previously from this source.
    \item Energy-dependent pulse profile: The pulse morphology showed a strong dependence on both photon energy and source intensity. The pulse shape evolves from simple single-peaked shapes at low energies to complex multi-peaked structures at intermediate energies and reverting to simpler morphologies at higher energies.
    \item Flares and emission geometry: Pronounced differences between flare and persistent pulse profiles indicate rapid changes in the emission geometry on short timescales. In particular, the disappearance of the secondary pulse peak and the increase in pulsed fraction during flares provide new evidence for transient accretion behaviour in \src. These flare-induced changes likely reflect magnetospheric instabilities or inhomogeneous mass inflow phenomena that have not been observed previously in this source.
    \item Soft excess and 1 keV line emission: Broadband spectral analysis revealed a soft thermal excess at $\sim$0.1 keV and a prominent emission feature near 1 keV, most likely originating from reprocessed emission in the inner accretion disc and contributions from Ne K and Fe L fluorescence, respectively. These spectral features were not reported in the previous study by \cite{Yang25}. We report the detection of the Ne K and Fe line complex for the first time in this source, enabled by the high sensitivity of the \nicer\ data. 
    \item Spin evolution and accretion torque: Phase-connected timing analysis over the \astrosat\ observation revealed a clear accretion-driven spin-up, with a measured frequency derivative of $\dot{\nu} \sim 2.4 \times 10^{-11}$ Hz s$^{-1}$. In addition, we also detected evidence for a decreasing spin-up rate as the outburst progressed, likely reflecting a gradual reduction in the mass accretion rate and corresponding accretion torque.
\end{enumerate}

Overall, our findings demonstrate that \src\ exhibits complex and rapidly evolving accretion dynamics during outburst, with both the pulsar’s spin and emission geometry responding sensitively to changes in the mass accretion rate. The combination of \astrosat\ and \nicer\ data provided complementary strengths. \astrosat\ offered broadband spectral coverage, while \nicer\ delivered high-cadence temporal sampling near the outburst peak and an independent timing baseline, consistency with those of \citet{Yang25} reinforces the reliability of our results. Moreover, the clear decline in $\dot{\nu}$ with decreasing flux provides direct observational evidence of torque–accretion rate coupling, offering new constraints on magnetospheric accretion processes and valuable insights into the coupling between accretion dynamics, pulse emission, and spin evolution in BeXRBs.

BeXRBs in the Magellanic Clouds offer unique advantages for such studies, as they lie at well-determined distances and are relatively unobscured by interstellar dust, unlike most Galactic BeXRBs, which often suffer from distance uncertainties and high extinction. \src, located in the LMC, has a well-constrained distance of 50 kpc with a smaller relative distance uncertainty of 2.2\% \citep{Pietrzynski13}. This minimizes uncertainty in the luminosity estimate and makes \src\ an ideal laboratory to study spin evolution, accretion torque, and emission properties of accreting pulsars, especially near-Eddington accretion rates.

\begin{acknowledgements}
      Data from the ToO phase of \astrosat{} observation were used in this study, obtained from the Indian Space Science Data Centre (ISSDC), and NICER data obtained from the High Energy Astrophysics Science Archive Research Center (HEASARC). We thank the LAXPC Payload Operation Center (POC) and the SXT POC at TIFR, Mumbai, for providing the necessary software tools. We thank the \nicer\ SOC Team for making the ToO observations possible. RS would like to thank Georgios Vasilopoulos for checking the consistency of the spin evolution obtained with \nicer. AB acknowledges the financial support from SERB (SB/SRS/2022-23/124/PS) and is grateful to the Royal Society, United Kingdom.
\end{acknowledgements}

\bibliographystyle{aa_url} 
\bibliography{ref} 

@ARTICLE{Sharma25-1626,
       author = {{Sharma}, Rahul and {Jain}, Chetana and {Paul}, Biswajit and {Beri}, Aru},
        title = "{Sidebands to mHz QPOs in 4U 1626-67 in the second spin-down state}",
      journal = {\mnras},
         year = 2025,
        month = apr,
       volume = {538},
       number = {2},
        pages = {1046-1054},
          doi = {10.1093/mnras/staf379},
archivePrefix = {arXiv},
       eprint = {2503.05444},
 primaryClass = {astro-ph.HE},
       adsurl = {https://ui.adsabs.harvard.edu/abs/2025MNRAS.538.1046S}
}

@ARTICLE{Sharma24,
       author = {{Sharma}, Rahul and {Mandal}, Manoj and {Pal}, Sabyasachi and {Paul}, Biswajit and {Jaisawal}, G.~K. and {Ratheesh}, Ajay},
        title = "{Probing the energy and luminosity-dependent spectro-timing properties of RX J0440.9+4431 with AstroSat}",
      journal = {\mnras},
         year = 2024,
        month = oct,
        pages = {1028–1042},
       volume = 534,
          doi = {10.1093/mnras/stae2175},
archivePrefix = {arXiv},
       eprint = {2409.11121},
 primaryClass = {astro-ph.HE},
       adsurl = {https://ui.adsabs.harvard.edu/abs/2024MNRAS.tmp.2127S}
}

@ARTICLE{Sharma24atel,
       author = {{Sharma}, Rahul and {Gendreau}, Keith and {Arzoumanian}, Zaven and {Ferrara}, Elizabeth C. and {Ray}, Paul S. and {Sanna}, Andrea},
        title = "{NICER confirms outburst of RX J0520.5-6932 and detects coherent pulsations}",
      journal = {The Astronomer's Telegram},
         year = 2024,
        month = apr,
       volume = {16569},
        pages = {1},
       adsurl = {https://ui.adsabs.harvard.edu/abs/2024ATel16569....1S}
}

@ARTICLE{Sharma23-lmc,
       author = {{Sharma}, Rahul and {Jain}, Chetana and {Rikame}, Ketan and {Paul}, Biswajit},
        title = "{Broad-band mHz QPOs and spectral study of LMC X-4 with AstroSat}",
      journal = {\mnras},
         year = 2023,
        month = feb,
       volume = {519},
       number = {2},
        pages = {1764-1770},
          doi = {10.1093/mnras/stac3572},
archivePrefix = {arXiv},
       eprint = {2212.01003},
 primaryClass = {astro-ph.HE},
       adsurl = {https://ui.adsabs.harvard.edu/abs/2023MNRAS.519.1764S}
}

@ARTICLE{Kumar2025,
       author = {{Kumar}, Manish and {Sharma}, Rahul and {Paul}, Biswajit and {Rana}, Vikram},
        title = "{Spectral analysis of ultraluminous X-ray pulsars with models of X-ray pulsars}",
      journal = {\mnras},
         year = 2025,
        month = jan,
       volume = {536},
       number = {1},
        pages = {340-349},
          doi = {10.1093/mnras/stae2558},
       adsurl = {https://ui.adsabs.harvard.edu/abs/2025MNRAS.536..340K}
}

@ARTICLE{Sharma2023-1626,
       author = {{Sharma}, Rahul and {Jain}, Chetana and {Paul}, Biswajit},
        title = "{4U 1626-67 returns to spin-down: timing features toe the line}",
      journal = {\mnras},
         year = 2023,
        month = nov,
       volume = {526},
       number = {1},
        pages = {L35-L40},
          doi = {10.1093/mnrasl/slad110},
archivePrefix = {arXiv},
       eprint = {2307.16599},
 primaryClass = {astro-ph.HE},
       adsurl = {https://ui.adsabs.harvard.edu/abs/2023MNRAS.526L..35S}
}

@ARTICLE{Wilms,
       author = {{Wilms}, J. and {Allen}, A. and {McCray}, R.},
        title = "{On the Absorption of X-Rays in the Interstellar Medium}",
      journal = {\apj},
         year = 2000,
        month = oct,
       volume = {542},
       number = {2},
        pages = {914-924},
          doi = {10.1086/317016},
archivePrefix = {arXiv},
       eprint = {astro-ph/0008425},
 primaryClass = {astro-ph},
       adsurl = {https://ui.adsabs.harvard.edu/abs/2000ApJ...542..914W}
}

@ARTICLE{Bonanos09,
       author = {{Bonanos}, A.~Z. and {Massa}, D.~L. and {Sewilo}, M. and {Lennon}, D.~J. and {Panagia}, N. and {Smith}, L.~J. and {Meixner}, M. and {Babler}, B.~L. and {Bracker}, S. and {Meade}, M.~R. and {Gordon}, K.~D. and {Hora}, J.~L. and {Indebetouw}, R. and {Whitney}, B.~A.},
        title = "{Spitzer SAGE Infrared Photometry of Massive Stars in the Large Magellanic Cloud}",
      journal = {\aj},
         year = 2009,
        month = oct,
       volume = {138},
       number = {4},
        pages = {1003-1021},
          doi = {10.1088/0004-6256/138/4/1003},
archivePrefix = {arXiv},
       eprint = {0905.1328},
 primaryClass = {astro-ph.SR},
       adsurl = {https://ui.adsabs.harvard.edu/abs/2009AJ....138.1003B}
}

@ARTICLE{Zhang24,
       author = {{Zhang}, Y.~J. and {Wang}, C.~Y. and {Liu}, Y. and {Xu}, X.~P. and {Ling}, Z.~X. and {Zhang}, C. and {Chen}, W. and {Cheng}, H.~Q. and {Cui}, C.~Z. and {Fan}, D.~W. and {Hu}, H.~B. and {Hu}, J.~W. and {Huang}, M.~H. and {Jin}, C.~C. and {Li}, D.~Y. and {Lian}, T.~Y. and {Liu}, H.~Y. and {Liu}, M.~J. and {Lv}, Z.~Z. and {Mao}, X. and {Pan}, H.~W. and {Pan}, X. and {Sun}, H. and {Wang}, W.~X. and {Wang}, Y.~L. and {Wu}, Q.~Y. and {Xu}, Y.~F. and {Yang}, H.~N. and {Zhang}, M. and {Zhang}, W.~D. and {Zhang}, W.~J. and {Zhao}, D.~H. and {Li}, D.~M. and {Li}, Q.~X. and {Yuan}, W.},
        title = "{LEIA detection of the brightening of the Be X-ray binary RX J0520.5-6932}",
      journal = {The Astronomer's Telegram},
         year = 2024,
        month = apr,
       volume = {16571},
        pages = {1},
       adsurl = {https://ui.adsabs.harvard.edu/abs/2024ATel16571....1Z}
}

@ARTICLE{Semena24,
       author = {{Semena}, A.~N. and {Mereminskiy}, I.~A. and {Lutovinov}, A.~A. and {Molkov}, S.~V. and {Tkachenko}, A. Yu. and {Arefiev}, V.~A.},
        title = "{SRG/ART-XC detection of a possible new giant outburst of RX J0520.5-6932}",
      journal = {The Astronomer's Telegram},
         year = 2024,
        month = mar,
       volume = {16562},
        pages = {1},
       adsurl = {https://ui.adsabs.harvard.edu/abs/2024ATel16562....1S}
}

@ARTICLE{Tendulkar14,
       author = {{Tendulkar}, Shriharsh P. and {F{\"u}rst}, Felix and {Pottschmidt}, Katja and {Bachetti}, Matteo and {Bhalerao}, Varun B. and {Boggs}, Steven E. and {Christensen}, Finn E. and {Craig}, William W. and {Hailey}, Charles A. and {Harrison}, Fiona A. and {Stern}, Daniel and {Tomsick}, John A. and {Walton}, Dominic J. and {Zhang}, William},
        title = "{NuSTAR Discovery of a Cyclotron Line in the Be/X-Ray Binary RX J0520.5-6932 during Outburst}",
      journal = {\apj},
         year = 2014,
        month = nov,
       volume = {795},
       number = {2},
          eid = {154},
        pages = {154},
          doi = {10.1088/0004-637X/795/2/154},
archivePrefix = {arXiv},
       eprint = {1409.5035},
 primaryClass = {astro-ph.HE},
       adsurl = {https://ui.adsabs.harvard.edu/abs/2014ApJ...795..154T}
}

@ARTICLE{Kuehnel14,
       author = {{Kuehnel}, M. and {Finger}, M.~H. and {Fuerst}, F. and {Pottschmidt}, K. and {Haberl}, F. and {Wilms}, J.},
        title = "{Orbital parameters and spin evolution of RX J0520.5-6932}",
      journal = {The Astronomer's Telegram},
         year = 2014,
        month = feb,
       volume = {5856},
        pages = {1},
       adsurl = {https://ui.adsabs.harvard.edu/abs/2014ATel.5856....1K}
}

@ARTICLE{Vasilopoulos14b,
       author = {{Vasilopoulos}, G. and {Haberl}, F. and {Sturm}, R. and {Maggi}, P. and {Udalski}, A.},
        title = "{Spectral and temporal properties of RX J0520.5-6932 (LXP 8.04) during a type-I outburst}",
      journal = {\aap},
         year = 2014,
        month = jul,
       volume = {567},
          eid = {A129},
        pages = {A129},
          doi = {10.1051/0004-6361/201423934},
archivePrefix = {arXiv},
       eprint = {1405.7312},
 primaryClass = {astro-ph.HE},
       adsurl = {https://ui.adsabs.harvard.edu/abs/2014A&A...567A.129V}
}

@ARTICLE{Vasilopoulos14a,
       author = {{Vasilopoulos}, G. and {Sturm}, R. and {Maggi}, P. and {Haberl}, F.},
        title = "{The X-ray outburst of RX J0520.5-6932 is reaching the Eddington luminosity}",
      journal = {The Astronomer's Telegram},
         year = 2014,
        month = jan,
       volume = {5760},
        pages = {1},
       adsurl = {https://ui.adsabs.harvard.edu/abs/2014ATel.5760....1V}
}

@ARTICLE{Vasilopoulos16,
       author = {{Vasilopoulos}, G. and {Haberl}, F. and {Delvaux}, C. and {Sturm}, R. and {Udalski}, A.},
        title = "{Multi-wavelength properties of IGR J05007-7047 (LXP 38.55) and identification as a Be X-ray binary pulsar in the LMC}",
      journal = {\mnras},
         year = 2016,
        month = sep,
       volume = {461},
       number = {2},
        pages = {1875-1884},
          doi = {10.1093/mnras/stw1408},
archivePrefix = {arXiv},
       eprint = {1606.04827},
 primaryClass = {astro-ph.HE},
       adsurl = {https://ui.adsabs.harvard.edu/abs/2016MNRAS.461.1875V}
}

@ARTICLE{DLmap90,
       author = {{Dickey}, John M. and {Lockman}, Felix J.},
        title = "{H I in the galaxy.}",
      journal = {\araa},
         year = 1990,
        month = jan,
       volume = {28},
        pages = {215-261},
          doi = {10.1146/annurev.aa.28.090190.001243},
       adsurl = {https://ui.adsabs.harvard.edu/abs/1990ARA&A..28..215D}
}

@ARTICLE{Rolleston02,
       author = {{Rolleston}, W.~R.~J. and {Trundle}, C. and {Dufton}, P.~L.},
        title = "{The present-day chemical composition of the LMC}",
      journal = {\aap},
         year = 2002,
        month = dec,
       volume = {396},
        pages = {53-64},
          doi = {10.1051/0004-6361:20021088},
       adsurl = {https://ui.adsabs.harvard.edu/abs/2002A&A...396...53R}
}

@ARTICLE{Edge04,
       author = {{Edge}, W.~R.~T. and {Coe}, M.~J. and {Galache}, J.~L. and {Hill}, A.~B.},
        title = "{A major outburst from the X-ray binary RX J0520.5-6932}",
      journal = {\mnras},
         year = 2004,
        month = apr,
       volume = {349},
       number = {4},
        pages = {1361-1364},
          doi = {10.1111/j.1365-2966.2004.07606.x},
archivePrefix = {arXiv},
       eprint = {astro-ph/0401537},
 primaryClass = {astro-ph},
       adsurl = {https://ui.adsabs.harvard.edu/abs/2004MNRAS.349.1361E}
}

@ARTICLE{Schmidtke94,
       author = {{Schmidtke}, P.~C. and {Cowley}, A.~P. and {Frattare}, L.~M. and {McGrath}, T.~K. and {Hutchings}, J.~B. and {Crampton}, D.},
        title = "{LMC Stellar X-Ray Sources Observed with ROSAT: I. X-Ray Data and Search for Optical Counterparts}",
      journal = {\pasp},
         year = 1994,
        month = aug,
       volume = {106},
        pages = {843},
          doi = {10.1086/133452},
       adsurl = {https://ui.adsabs.harvard.edu/abs/1994PASP..106..843S}
}

@ARTICLE{Karaferias23,
       author = {{Karaferias}, A.~S. and {Vasilopoulos}, G. and {Petropoulou}, M. and {Jenke}, P.~A. and {Wilson-Hodge}, C.~A. and {Malacaria}, C.},
        title = "{A Bayesian approach for torque modelling of BeXRB pulsars with application to super-Eddington accretors}",
      journal = {\mnras},
         year = 2023,
        month = mar,
       volume = {520},
       number = {1},
        pages = {281-299},
          doi = {10.1093/mnras/stac3208},
archivePrefix = {arXiv},
       eprint = {2211.16079},
 primaryClass = {astro-ph.HE},
       adsurl = {https://ui.adsabs.harvard.edu/abs/2023MNRAS.520..281K}
}

@ARTICLE{Agrawal2006,
       author = {{Agrawal}, P.~C.},
        title = "{A broad spectral band Indian Astronomy satellite {\textquoteleft}Astrosat{\textquoteright}}",
      journal = {Advances in Space Research},
         year = 2006,
        month = jan,
       volume = {38},
       number = {12},
        pages = {2989-2994},
          doi = {10.1016/j.asr.2006.03.038},
       adsurl = {https://ui.adsabs.harvard.edu/abs/2006AdSpR..38.2989A}
}

@ARTICLE{Antia2017,
       author = {{Antia}, H.~M. and {Yadav}, J.~S. and {Agrawal}, P.~C. and {Verdhan Chauhan}, Jai and {Manchanda}, R.~K. and {Chitnis}, Varsha and {Paul}, Biswajit and {Dedhia}, Dhiraj and {Shah}, Parag and {Gujar}, V.~M. and {Katoch}, Tilak and {Kurhade}, V.~N. and {Madhwani}, Pankaj and {Manojkumar}, T.~K. and {Nikam}, V.~A. and {Pandya}, A.~S. and {Parmar}, J.~V. and {Pawar}, D.~M. and {Pahari}, Mayukh and {Misra}, Ranjeev and {Navalgund}, K.~H. and {Pandiyan}, R. and {Sharma}, K.~S. and {Subbarao}, K.},
        title = "{Calibration of the Large Area X-Ray Proportional Counter (LAXPC) Instrument on board AstroSat}",
      journal = {\apjs},
         year = 2017,
        month = jul,
       volume = {231},
       number = {1},
          eid = {10},
        pages = {10},
          doi = {10.3847/1538-4365/aa7a0e},
archivePrefix = {arXiv},
       eprint = {1702.08624},
 primaryClass = {astro-ph.IM},
       adsurl = {https://ui.adsabs.harvard.edu/abs/2017ApJS..231...10A}
}

@INPROCEEDINGS{Arnaud,
       author = {{Arnaud}, K.~A.},
        title = "{XSPEC: The First Ten Years}",
    booktitle = {Astronomical Data Analysis Software and Systems V},
         year = 1996,
       editor = {{Jacoby}, George H. and {Barnes}, Jeannette},
       series = {Astronomical Society of the Pacific Conference Series},
       volume = {101},
        month = jan,
        pages = {17},
       adsurl = {https://ui.adsabs.harvard.edu/abs/1996ASPC..101...17A}
}

@ARTICLE{Beri2017,
       author = {{Beri}, Aru and {Paul}, Biswajit},
        title = "{Post-flare formation of the accretion stream and a dip in pulse profiles of LMC X-4}",
      journal = {\na},
         year = 2017,
        month = oct,
       volume = {56},
        pages = {94-101},
          doi = {10.1016/j.newast.2017.05.001},
archivePrefix = {arXiv},
       eprint = {1705.05205},
 primaryClass = {astro-ph.HE},
       adsurl = {https://ui.adsabs.harvard.edu/abs/2017NewA...56...94B}
}

@ARTICLE{Boldin2013,
       author = {{Boldin}, P.~A. and {Tsygankov}, S.~S. and {Lutovinov}, A.~A.},
        title = "{On timing and spectral characteristics of the X-ray pulsar 4U 0115+63: Evolution of the pulsation period and the cyclotron line energy}",
      journal = {Astronomy Letters},
         year = 2013,
        month = jun,
       volume = {39},
       number = {6},
        pages = {375-388},
          doi = {10.1134/S1063773713060029},
archivePrefix = {arXiv},
       eprint = {1305.6785},
 primaryClass = {astro-ph.HE},
       adsurl = {https://ui.adsabs.harvard.edu/abs/2013AstL...39..375B}
}

@ARTICLE{Hickox2004,
       author = {{Hickox}, Ryan C. and {Narayan}, Ramesh and {Kallman}, Timothy R.},
        title = "{Origin of the Soft Excess in X-Ray Pulsars}",
      journal = {\apj},
         year = 2004,
        month = oct,
       volume = {614},
       number = {2},
        pages = {881-896},
          doi = {10.1086/423928},
archivePrefix = {arXiv},
       eprint = {astro-ph/0407115},
 primaryClass = {astro-ph},
       adsurl = {https://ui.adsabs.harvard.edu/abs/2004ApJ...614..881H}
}

@ARTICLE{Moon01,
       author = {{Moon}, Dae-Sik and {Eikenberry}, Stephen S.},
        title = "{Large X-Ray Flares from LMC X-4: Discovery of Millihertz Quasi-periodic Oscillations and Quasi-periodic Oscillation-Modulated Pulsations}",
      journal = {\apjl},
         year = 2001,
        month = mar,
       volume = {549},
       number = {2},
        pages = {L225-L228},
          doi = {10.1086/319160},
archivePrefix = {arXiv},
       eprint = {astro-ph/0101393},
 primaryClass = {astro-ph},
       adsurl = {https://ui.adsabs.harvard.edu/abs/2001ApJ...549L.225M}
}

@ARTICLE{Naik2004,
       author = {{Naik}, S. and {Paul}, B.},
        title = "{Timing and Spectral Studies of LMC X-4 in High and Low States with BeppoSAX: Detection of Pulsations in the Soft Spectral Component}",
      journal = {\apj},
         year = 2004,
        month = jan,
       volume = {600},
       number = {1},
        pages = {351-357},
          doi = {10.1086/379803},
archivePrefix = {arXiv},
       eprint = {astro-ph/0309431},
 primaryClass = {astro-ph},
       adsurl = {https://ui.adsabs.harvard.edu/abs/2004ApJ...600..351N}
}

@ARTICLE{Paul2002,
       author = {{Paul}, B. and {Nagase}, F. and {Endo}, T. and {Dotani}, T. and {Yokogawa}, J. and {Nishiuchi}, M.},
        title = "{Nature of the Soft Spectral Component in the X-Ray Pulsars SMC X-1 and LMC X-4}",
      journal = {\apj},
         year = 2002,
        month = nov,
       volume = {579},
       number = {1},
        pages = {411-421},
          doi = {10.1086/342701},
archivePrefix = {arXiv},
       eprint = {astro-ph/0207341},
 primaryClass = {astro-ph},
       adsurl = {https://ui.adsabs.harvard.edu/abs/2002ApJ...579..411P}
}

@ARTICLE{Pietrzynski13,
       author = {{Pietrzy{\'n}ski}, G. and {Graczyk}, D. and {Gieren}, W. and {Thompson}, I.~B. and {Pilecki}, B. and {Udalski}, A. and {Soszy{\'n}ski}, I. and {Koz{\l}owski}, S. and {Konorski}, P. and {Suchomska}, K. and {Bono}, G. and {Moroni}, P.~G. Prada and {Villanova}, S. and {Nardetto}, N. and {Bresolin}, F. and {Kudritzki}, R.~P. and {Storm}, J. and {Gallenne}, A. and {Smolec}, R. and {Minniti}, D. and {Kubiak}, M. and {Szyma{\'n}ski}, M.~K. and {Poleski}, R. and {Wyrzykowski}, {\L}. and {Ulaczyk}, K. and {Pietrukowicz}, P. and {G{\'o}rski}, M. and {Karczmarek}, P.},
        title = "{An eclipsing-binary distance to the Large Magellanic Cloud accurate to two per cent}",
      journal = {\nat},
         year = 2013,
        month = mar,
       volume = {495},
       number = {7439},
        pages = {76-79},
          doi = {10.1038/nature11878},
archivePrefix = {arXiv},
       eprint = {1303.2063},
 primaryClass = {astro-ph.GA},
       adsurl = {https://ui.adsabs.harvard.edu/abs/2013Natur.495...76P}
}

@ARTICLE{Shtykovsky2018,
       author = {{Shtykovsky}, A.~E. and {Arefiev}, V.~A. and {Lutovinov}, A.~A. and {Molkov}, S.~V.},
        title = "{Peculiarities of Super-Eddington Flares from the X-ray Pulsar LMC X-4 Based on NuSTAR Data}",
      journal = {Astronomy Letters},
         year = 2018,
        month = mar,
       volume = {44},
       number = {3},
        pages = {149-161},
          doi = {10.1134/S1063773718030015},
archivePrefix = {arXiv},
       eprint = {1712.05322},
 primaryClass = {astro-ph.HE},
       adsurl = {https://ui.adsabs.harvard.edu/abs/2018AstL...44..149S}
}

@ARTICLE{Singh2017,
       author = {{Singh}, K.~P. and {Stewart}, G.~C. and {Westergaard}, N.~J. and {Bhattacharayya}, S. and {Chandra}, S. and {Chitnis}, V.~R. and {Dewangan}, G.~C. and {Kothare}, A.~T. and {Mirza}, I.~M. and {Mukerjee}, K. and {Navalkar}, V. and {Shah}, H. and {Abbey}, A.~F. and {Beardmore}, A.~P. and {Kotak}, S. and {Kamble}, N. and {Vishwakarama}, S. and {Pathare}, D.~P. and {Risbud}, V.~M. and {Koyande}, J.~P. and {Stevenson}, T. and {Bicknell}, C. and {Crawford}, T. and {Hansford}, G. and {Peters}, G. and {Sykes}, J. and {Agarwal}, P. and {Sebastian}, M. and {Rajarajan}, A. and {Nagesh}, G. and {Narendra}, S. and {Ramesh}, M. and {Rai}, R. and {Navalgund}, K.~H. and {Sarma}, K.~S. and {Pandiyan}, R. and {Subbarao}, K. and {Gupta}, T. and {Thakkar}, N. and {Singh}, A.~K. and {Bajpai}, A.},
        title = "{Soft X-ray Focusing Telescope Aboard AstroSat: Design, Characteristics and Performance}",
      journal = {Journal of Astrophysics and Astronomy},
         year = 2017,
        month = jun,
       volume = {38},
       number = {2},
          eid = {29},
        pages = {29},
          doi = {10.1007/s12036-017-9448-7},
       adsurl = {https://ui.adsabs.harvard.edu/abs/2017JApA...38...29S}
}

@INPROCEEDINGS{Singh2014,
       author = {{Singh}, Kulinder Pal and {Tandon}, S.~N. and {Agrawal}, P.~C. and {Antia}, H.~M. and {Manchanda}, R.~K. and {Yadav}, J.~S. and {Seetha}, S. and {Ramadevi}, M.~C. and {Rao}, A.~R. and {Bhattacharya}, D. and {Paul}, B. and {Sreekumar}, P. and {Bhattacharyya}, S. and {Stewart}, G.~C. and {Hutchings}, J. and {Annapurni}, S.~A. and {Ghosh}, S.~K. and {Murthy}, J. and {Pati}, A. and {Rao}, N.~K. and {Stalin}, C.~S. and {Girish}, V. and {Sankarasubramanian}, K. and {Vadawale}, S. and {Bhalerao}, V.~B. and {Dewangan}, G.~C. and {Dedhia}, D.~K. and {Hingar}, M.~K. and {Katoch}, T.~B. and {Kothare}, A.~T. and {Mirza}, I. and {Mukerjee}, K. and {Shah}, H. and {Shah}, P. and {Mohan}, R. and {Sangal}, A.~K. and {Nagabhusana}, S. and {Sriram}, S. and {Malkar}, J.~P. and {Sreekumar}, S. and {Abbey}, A.~F. and {Hansford}, G.~M. and {Beardmore}, A.~P. and {Sharma}, M.~R. and {Murthy}, S. and {Kulkarni}, R. and {Meena}, G. and {Babu}, V.~C. and {Postma}, J.},
        title = "{ASTROSAT mission}",
    booktitle = {Space Telescopes and Instrumentation 2014: Ultraviolet to Gamma Ray},
         year = 2014,
       editor = {{Takahashi}, Tadayuki and {den Herder}, Jan-Willem A. and {Bautz}, Mark},
       series = {Society of Photo-Optical Instrumentation Engineers (SPIE) Conference Series},
       volume = {9144},
        month = jul,
          eid = {91441S},
        pages = {91441S},
          doi = {10.1117/12.2062667},
       adsurl = {https://ui.adsabs.harvard.edu/abs/2014SPIE.9144E..1SS}
}

@INPROCEEDINGS{Singh2016,
       author = {{Singh}, Kulinder Pal and {Stewart}, Gordon C. and {Chandra}, Sunil and {Mukerjee}, Kallol and {Kotak}, Sanket and {Beardmore}, Andy P. and {Chitnis}, Varsha and {Dewangan}, Gulab C. and {Bhattacharyya}, Sudip and {Mirza}, Irfan and {Kamble}, Nilima and {Navalkar}, Vinita and {Shah}, Harshit and {Vishwakarma}, S. and {Koyande}, J.},
        title = "{In-orbit performance of SXT aboard AstroSat}",
    booktitle = {Space Telescopes and Instrumentation 2016: Ultraviolet to Gamma Ray},
         year = 2016,
       editor = {{den Herder}, Jan-Willem A. and {Takahashi}, Tadayuki and {Bautz}, Marshall},
       series = {Society of Photo-Optical Instrumentation Engineers (SPIE) Conference Series},
       volume = {9905},
        month = jul,
          eid = {99051E},
        pages = {99051E},
          doi = {10.1117/12.2235309},
       adsurl = {https://ui.adsabs.harvard.edu/abs/2016SPIE.9905E..1ES}
}

@INPROCEEDINGS{Yadav2016,
       author = {{Yadav}, J.~S. and {Agrawal}, P.~C. and {Antia}, H.~M. and {Chauhan}, Jai Verdhan and {Dedhia}, Dhiraj and {Katoch}, Tilak and {Madhwani}, P. and {Manchanda}, R.~K. and {Misra}, Ranjeev and {Pahari}, Mayukh and {Paul}, B. and {Shah}, Parag},
        title = "{Large Area X-ray Proportional Counter (LAXPC) instrument onboard ASTROSAT}",
    booktitle = {Space Telescopes and Instrumentation 2016: Ultraviolet to Gamma Ray},
         year = 2016,
       editor = {{den Herder}, Jan-Willem A. and {Takahashi}, Tadayuki and {Bautz}, Marshall},
       series = {Society of Photo-Optical Instrumentation Engineers (SPIE) Conference Series},
       volume = {9905},
        month = jul,
          eid = {99051D},
        pages = {99051D},
          doi = {10.1117/12.2231857},
       adsurl = {https://ui.adsabs.harvard.edu/abs/2016SPIE.9905E..1DY}
}

@ARTICLE{James2011,
       author = {{James}, Marykutty and {Paul}, Biswajit and {Devasia}, Jincy and {Indulekha}, Kavila},
        title = "{Flares, broadening of the pulse-frequency peak and quasi-periodic oscillations in the transient X-ray pulsar 4U 1901+03}",
      journal = {\mnras},
         year = 2011,
        month = jan,
       volume = {410},
       number = {3},
        pages = {1489-1495},
          doi = {10.1111/j.1365-2966.2010.17543.x},
archivePrefix = {arXiv},
       eprint = {1008.2815},
 primaryClass = {astro-ph.HE},
       adsurl = {https://ui.adsabs.harvard.edu/abs/2011MNRAS.410.1489J}
}

@ARTICLE{Yang25,
       author = {{Yang}, H.~N. and {Maitra}, C. and {Vasilopoulos}, G. and {Haberl}, F. and {Jenke}, P.~A. and {Karaferias}, A.~S. and {Sharma}, R. and {Beri}, A. and {Ji}, L. and {Jin}, C. and {Yuan}, W. and {Zhang}, Y.~J. and {Wang}, C.~Y. and {Xu}, X.~P. and {Liu}, Y. and {Zhang}, W.~D. and {Zhang}, C. and {Ling}, Z.~X. and {Liu}, H.~Y. and {Cheng}, H.~Q. and {Pan}, H.~W.},
        title = "{Broad-band study of the Be X-ray binary RX J0520.5-6932 during its outburst in 2024}",
      journal = {\mnras},
         year = 2025,
        month = jan,
       volume = {536},
       number = {2},
        pages = {1357-1373},
          doi = {10.1093/mnras/stae2676},
archivePrefix = {arXiv},
       eprint = {2412.00960},
 primaryClass = {astro-ph.HE},
       adsurl = {https://ui.adsabs.harvard.edu/abs/2025MNRAS.536.1357Y}
}

@ARTICLE{Misra21,
       author = {{Misra}, Ranjeev and {Roy}, Jayashree and {Yadav}, J.~S.},
        title = "{An alternative scheme to estimate AstroSat/LAXPC background for faint sources}",
      journal = {Journal of Astrophysics and Astronomy},
         year = 2021,
        month = oct,
       volume = {42},
       number = {2},
          eid = {55},
        pages = {55},
          doi = {10.1007/s12036-021-09734-2},
archivePrefix = {arXiv},
       eprint = {2102.06402},
 primaryClass = {astro-ph.IM},
       adsurl = {https://ui.adsabs.harvard.edu/abs/2021JApA...42...55M}
}

@ARTICLE{Kaastra,
       author = {{Kaastra}, J.~S. and {Bleeker}, J.~A.~M.},
        title = "{Optimal binning of X-ray spectra and response matrix design}",
      journal = {\aap},
         year = 2016,
        month = mar,
       volume = {587},
          eid = {A151},
        pages = {A151},
          doi = {10.1051/0004-6361/201527395},
archivePrefix = {arXiv},
       eprint = {1601.05309},
 primaryClass = {astro-ph.IM},
       adsurl = {https://ui.adsabs.harvard.edu/abs/2016A&A...587A.151K}
}

@INPROCEEDINGS{nicer,
       author = {{Gendreau}, Keith C. and {Arzoumanian}, Zaven and {Adkins}, Phillip W. and {Albert}, Cheryl L. and {Anders}, John F. and {Aylward}, Andrew T. and {Baker}, Charles L. and {Balsamo}, Erin R. and {Bamford}, William A. and {Benegalrao}, Suyog S. and {Berry}, Daniel L. and {Bhalwani}, Shiraz and {Black}, J. Kevin and {Blaurock}, Carl and {Bronke}, Ginger M. and {Brown}, Gary L. and {Budinoff}, Jason G. and {Cantwell}, Jeffrey D. and {Cazeau}, Thoniel and {Chen}, Philip T. and {Clement}, Thomas G. and {Colangelo}, Andrew T. and {Coleman}, Jerry S. and {Coopersmith}, Jonathan D. and {Dehaven}, William E. and {Doty}, John P. and {Egan}, Mark D. and {Enoto}, Teruaki and {Fan}, Terry W. and {Ferro}, Deneen M. and {Foster}, Richard and {Galassi}, Nicholas M. and {Gallo}, Luis D. and {Green}, Chris M. and {Grosh}, Dave and {Ha}, Kong Q. and {Hasouneh}, Monther A. and {Heefner}, Kristofer B. and {Hestnes}, Phyllis and {Hoge}, Lisa J. and {Jacobs}, Tawanda M. and {J{\o}rgensen}, John L. and {Kaiser}, Michael A. and {Kellogg}, James W. and {Kenyon}, Steven J. and {Koenecke}, Richard G. and {Kozon}, Robert P. and {LaMarr}, Beverly and {Lambertson}, Mike D. and {Larson}, Anne M. and {Lentine}, Steven and {Lewis}, Jesse H. and {Lilly}, Michael G. and {Liu}, Kuochia Alice and {Malonis}, Andrew and {Manthripragada}, Sridhar S. and {Markwardt}, Craig B. and {Matonak}, Bryan D. and {Mcginnis}, Isaac E. and {Miller}, Roger L. and {Mitchell}, Alissa L. and {Mitchell}, Jason W. and {Mohammed}, Jelila S. and {Monroe}, Charles A. and {Montt de Garcia}, Kristina M. and {Mul{\'e}}, Peter D. and {Nagao}, Louis T. and {Ngo}, Son N. and {Norris}, Eric D. and {Norwood}, Dwight A. and {Novotka}, Joseph and {Okajima}, Takashi and {Olsen}, Lawrence G. and {Onyeachu}, Chimaobi O. and {Orosco}, Henry Y. and {Peterson}, Jacqualine R. and {Pevear}, Kristina N. and {Pham}, Karen K. and {Pollard}, Sue E. and {Pope}, John S. and {Powers}, Daniel F. and {Powers}, Charles E. and {Price}, Samuel R. and {Prigozhin}, Gregory Y. and {Ramirez}, Julian B. and {Reid}, Winston J. and {Remillard}, Ronald A. and {Rogstad}, Eric M. and {Rosecrans}, Glenn P. and {Rowe}, John N. and {Sager}, Jennifer A. and {Sanders}, Claude A. and {Savadkin}, Bruce and {Saylor}, Maxine R. and {Schaeffer}, Alexander F. and {Schweiss}, Nancy S. and {Semper}, Sean R. and {Serlemitsos}, Peter J. and {Shackelford}, Larry V. and {Soong}, Yang and {Struebel}, Jonathan and {Vezie}, Michael L. and {Villasenor}, Joel S. and {Winternitz}, Luke B. and {Wofford}, George I. and {Wright}, Michael R. and {Yang}, Mike Y. and {Yu}, Wayne H.},
        title = "{The Neutron star Interior Composition Explorer (NICER): design and development}",
    booktitle = {Space Telescopes and Instrumentation 2016: Ultraviolet to Gamma Ray},
         year = 2016,
       editor = {{den Herder}, Jan-Willem A. and {Takahashi}, Tadayuki and {Bautz}, Marshall},
       series = {Society of Photo-Optical Instrumentation Engineers (SPIE) Conference Series},
       volume = {9905},
        month = jul,
          eid = {99051H},
        pages = {99051H},
          doi = {10.1117/12.2231304},
       adsurl = {https://ui.adsabs.harvard.edu/abs/2016SPIE.9905E..1HG}
}

@ARTICLE{Remillard22,
       author = {{Remillard}, Ronald A. and {Loewenstein}, Michael and {Steiner}, James F. and {Prigozhin}, Gregory Y. and {LaMarr}, Beverly and {Enoto}, Teruaki and {Gendreau}, Keith C. and {Arzoumanian}, Zaven and {Markwardt}, Craig and {Basak}, Arkadip and {Stevens}, Abigail L. and {Ray}, Paul S. and {Altamirano}, Diego and {Buisson}, Douglas J.~K.},
        title = "{An Empirical Background Model for the NICER X-Ray Timing Instrument}",
      journal = {\aj},
         year = 2022,
        month = mar,
       volume = {163},
       number = {3},
          eid = {130},
        pages = {130},
          doi = {10.3847/1538-3881/ac4ae6},
archivePrefix = {arXiv},
       eprint = {2105.09901},
 primaryClass = {astro-ph.IM},
       adsurl = {https://ui.adsabs.harvard.edu/abs/2022AJ....163..130R}
}

@ARTICLE{Reig11,
       author = {{Reig}, Pablo},
        title = "{Be/X-ray binaries}",
      journal = {\apss},
         year = 2011,
        month = mar,
       volume = {332},
       number = {1},
        pages = {1-29},
          doi = {10.1007/s10509-010-0575-8},
archivePrefix = {arXiv},
       eprint = {1101.5036},
 primaryClass = {astro-ph.HE},
       adsurl = {https://ui.adsabs.harvard.edu/abs/2011Ap&SS.332....1R}
}

@ARTICLE{Liu06,
       author = {{Liu}, Q.~Z. and {van Paradijs}, J. and {van den Heuvel}, E.~P.~J.},
        title = "{Catalogue of high-mass X-ray binaries in the Galaxy (4th edition)}",
      journal = {\aap},
         year = 2006,
        month = sep,
       volume = {455},
       number = {3},
        pages = {1165-1168},
          doi = {10.1051/0004-6361:20064987},
archivePrefix = {arXiv},
       eprint = {0707.0549},
 primaryClass = {astro-ph},
       adsurl = {https://ui.adsabs.harvard.edu/abs/2006A&A...455.1165L}
}

@ARTICLE{Basko76,
       author = {{Basko}, M.~M. and {Sunyaev}, R.~A.},
        title = "{The limiting luminosity of accreting neutron stars with magnetic fields.}",
      journal = {\mnras},
         year = 1976,
        month = may,
       volume = {175},
        pages = {395-417},
          doi = {10.1093/mnras/175.2.395},
       adsurl = {https://ui.adsabs.harvard.edu/abs/1976MNRAS.175..395B}
}

@ARTICLE{Becker12,
       author = {{Becker}, P.~A. and {Klochkov}, D. and {Sch{\"o}nherr}, G. and {Nishimura}, O. and {Ferrigno}, C. and {Caballero}, I. and {Kretschmar}, P. and {Wolff}, M.~T. and {Wilms}, J. and {Staubert}, R.},
        title = "{Spectral formation in accreting X-ray pulsars: bimodal variation of the cyclotron energy with luminosity}",
      journal = {\aap},
         year = 2012,
        month = aug,
       volume = {544},
          eid = {A123},
        pages = {A123},
          doi = {10.1051/0004-6361/201219065},
archivePrefix = {arXiv},
       eprint = {1205.5316},
 primaryClass = {astro-ph.HE},
       adsurl = {https://ui.adsabs.harvard.edu/abs/2012A&A...544A.123B}
}

@ARTICLE{Belloni02,
       author = {{Belloni}, Tomaso and {Psaltis}, Dimitrios and {van der Klis}, Michiel},
        title = "{A Unified Description of the Timing Features of Accreting X-Ray Binaries}",
      journal = {\apj},
         year = 2002,
        month = jun,
       volume = {572},
       number = {1},
        pages = {392-406},
          doi = {10.1086/340290},
archivePrefix = {arXiv},
       eprint = {astro-ph/0202213},
 primaryClass = {astro-ph},
       adsurl = {https://ui.adsabs.harvard.edu/abs/2002ApJ...572..392B}
}

@ARTICLE{Bozzo09,
       author = {{Bozzo}, E. and {Stella}, L. and {Vietri}, M. and {Ghosh}, P.},
        title = "{Can disk-magnetosphere interaction models and beat frequency models for quasi-periodic oscillation in accreting X-ray pulsars be reconciled?}",
      journal = {\aap},
         year = 2009,
        month = jan,
       volume = {493},
       number = {3},
        pages = {809-818},
          doi = {10.1051/0004-6361:200810658},
archivePrefix = {arXiv},
       eprint = {0811.0049},
 primaryClass = {astro-ph},
       adsurl = {https://ui.adsabs.harvard.edu/abs/2009A&A...493..809B}
}

@ARTICLE{Pringle72,
       author = {{Pringle}, J.~E. and {Rees}, M.~J.},
        title = "{Accretion Disc Models for Compact X-Ray Sources}",
      journal = {\aap},
         year = 1972,
        month = oct,
       volume = {21},
        pages = {1},
       adsurl = {https://ui.adsabs.harvard.edu/abs/1972A&A....21....1P}
}

@ARTICLE{Ghosh79-III,
       author = {{Ghosh}, P. and {Lamb}, F.~K.},
        title = "{Accretion by rotating magnetic neutron stars. III. Accretion torques and period changes in pulsating X-ray sources.}",
      journal = {\apj},
         year = 1979,
        month = nov,
       volume = {234},
        pages = {296-316},
          doi = {10.1086/157498},
       adsurl = {https://ui.adsabs.harvard.edu/abs/1979ApJ...234..296G}
}

@ARTICLE{Leahy87,
       author = {{Leahy}, D.~A.},
        title = "{Searches for pulsed emission - Improved determination of period and amplitude from epoch folding for sinusoidal signals}",
      journal = {\aap},
         year = 1987,
        month = jun,
       volume = {180},
       number = {1-2},
        pages = {275-277},
       adsurl = {https://ui.adsabs.harvard.edu/abs/1987A&A...180..275L}
}

@ARTICLE{Lutovinov09,
       author = {{Lutovinov}, A.~A. and {Tsygankov}, S.~S.},
        title = "{Timing characteristics of the hard X-ray emission from bright X-ray pulsars based on INTEGRAL data}",
      journal = {Astronomy Letters},
         year = 2009,
        month = jul,
       volume = {35},
       number = {7},
        pages = {433-456},
          doi = {10.1134/S1063773709070019},
archivePrefix = {arXiv},
       eprint = {0907.4288},
 primaryClass = {astro-ph.HE},
       adsurl = {https://ui.adsabs.harvard.edu/abs/2009AstL...35..433L}
}

@ARTICLE{Lyubarskii97,
       author = {{Lyubarskii}, Yu. E.},
        title = "{Flicker noise in accretion discs}",
      journal = {\mnras},
         year = 1997,
        month = dec,
       volume = {292},
       number = {3},
        pages = {679-685},
          doi = {10.1093/mnras/292.3.679},
       adsurl = {https://ui.adsabs.harvard.edu/abs/1997MNRAS.292..679L}
}

@ARTICLE{Malacaria20GBM,
       author = {{Malacaria}, C. and {Jenke}, P. and {Roberts}, O.~J. and {Wilson-Hodge}, C.~A. and {Cleveland}, W.~H. and {Mailyan}, B. and {GBM Accreting Pulsars Program Team}},
        title = "{The Ups and Downs of Accreting X-Ray Pulsars: Decade-long Observations with the Fermi Gamma-Ray Burst Monitor}",
      journal = {\apj},
         year = 2020,
        month = jun,
       volume = {896},
       number = {1},
          eid = {90},
        pages = {90},
          doi = {10.3847/1538-4357/ab855c},
archivePrefix = {arXiv},
       eprint = {2004.00051},
 primaryClass = {astro-ph.HE},
       adsurl = {https://ui.adsabs.harvard.edu/abs/2020ApJ...896...90M}
}

@ARTICLE{Mushtukov22,
       author = {{Mushtukov}, Alexander and {Tsygankov}, Sergey},
        title = "{Accreting strongly magnetised neutron stars: X-ray Pulsars}",
      journal = {arXiv e-prints},
         year = 2022,
        month = apr,
          eid = {arXiv:2204.14185},
        pages = {arXiv:2204.14185},
          doi = {10.48550/arXiv.2204.14185},
archivePrefix = {arXiv},
       eprint = {2204.14185},
 primaryClass = {astro-ph.HE},
       adsurl = {https://ui.adsabs.harvard.edu/abs/2022arXiv220414185M}
}

@ARTICLE{Mushtukov15,
       author = {{Mushtukov}, Alexander A. and {Suleimanov}, Valery F. and {Tsygankov}, Sergey S. and {Poutanen}, Juri},
        title = "{The critical accretion luminosity for magnetized neutron stars}",
      journal = {\mnras},
         year = 2015,
        month = feb,
       volume = {447},
       number = {2},
        pages = {1847-1856},
          doi = {10.1093/mnras/stu2484},
archivePrefix = {arXiv},
       eprint = {1409.6457},
 primaryClass = {astro-ph.HE},
       adsurl = {https://ui.adsabs.harvard.edu/abs/2015MNRAS.447.1847M}
}

@ARTICLE{Reig08,
       author = {{Reig}, P.},
        title = "{Rapid spectral and timing variability of Be/X-ray binaries during type ;II outbursts}",
      journal = {\aap},
         year = 2008,
        month = oct,
       volume = {489},
       number = {2},
        pages = {725-740},
          doi = {10.1051/0004-6361:200810021},
archivePrefix = {arXiv},
       eprint = {0807.4786},
 primaryClass = {astro-ph},
       adsurl = {https://ui.adsabs.harvard.edu/abs/2008A&A...489..725R}
}

@ARTICLE{Revnivtsev09,
       author = {{Revnivtsev}, M. and {Churazov}, E. and {Postnov}, K. and {Tsygankov}, S.},
        title = "{Quenching of the strong aperiodic accretion disk variability at the magnetospheric boundary}",
      journal = {\aap},
         year = 2009,
        month = dec,
       volume = {507},
       number = {3},
        pages = {1211-1215},
          doi = {10.1051/0004-6361/200912317},
archivePrefix = {arXiv},
       eprint = {0909.2996},
 primaryClass = {astro-ph.HE},
       adsurl = {https://ui.adsabs.harvard.edu/abs/2009A&A...507.1211R}
}

@ARTICLE{Wilson18,
       author = {{Wilson-Hodge}, Colleen A. and {Malacaria}, Christian and {Jenke}, Peter A. and {Jaisawal}, Gaurava K. and {Kerr}, Matthew and {Wolff}, Michael T. and {Arzoumanian}, Zaven and {Chakrabarty}, Deepto and {Doty}, John P. and {Gendreau}, Keith C. and {Guillot}, Sebastien and {Ho}, Wynn C.~G. and {LaMarr}, Beverly and {Markwardt}, Craig B. and {{\"O}zel}, Feryal and {Prigozhin}, Gregory Y. and {Ray}, Paul S. and {Ramos-Lerate}, Mercedes and {Remillard}, Ronald A. and {Strohmayer}, Tod E. and {Vezie}, Michael L. and {Wood}, Kent S. and {NICER Science Team}},
        title = "{NICER and Fermi GBM Observations of the First Galactic Ultraluminous X-Ray Pulsar Swift J0243.6+6124}",
      journal = {\apj},
         year = 2018,
        month = aug,
       volume = {863},
       number = {1},
          eid = {9},
        pages = {9},
          doi = {10.3847/1538-4357/aace60},
archivePrefix = {arXiv},
       eprint = {1806.10094},
 primaryClass = {astro-ph.HE},
       adsurl = {https://ui.adsabs.harvard.edu/abs/2018ApJ...863....9W}
}

@ARTICLE{Ferrigno23,
       author = {{Ferrigno}, Carlo and {D'A{\`\i}}, Antonino and {Ambrosi}, Elena},
        title = "{Energy-resolved pulse profiles of accreting pulsars: Diagnostic tools for spectral features}",
      journal = {\aap},
         year = 2023,
        month = sep,
       volume = {677},
          eid = {A103},
        pages = {A103},
          doi = {10.1051/0004-6361/202347062},
archivePrefix = {arXiv},
       eprint = {2308.03395},
 primaryClass = {astro-ph.HE},
       adsurl = {https://ui.adsabs.harvard.edu/abs/2023A&A...677A.103F}
}

@ARTICLE{Coburn02,
       author = {{Coburn}, W. and {Heindl}, W.~A. and {Rothschild}, R.~E. and {Gruber}, D.~E. and {Kreykenbohm}, I. and {Wilms}, J. and {Kretschmar}, P. and {Staubert}, R.},
        title = "{Magnetic Fields of Accreting X-Ray Pulsars with the Rossi X-Ray Timing Explorer}",
      journal = {\apj},
         year = 2002,
        month = nov,
       volume = {580},
       number = {1},
        pages = {394-412},
          doi = {10.1086/343033},
archivePrefix = {arXiv},
       eprint = {astro-ph/0207325},
 primaryClass = {astro-ph},
       adsurl = {https://ui.adsabs.harvard.edu/abs/2002ApJ...580..394C}
}

@ARTICLE{Maitra18,
       author = {{Maitra}, C. and {Paul}, B. and {Haberl}, F. and {Vasilopoulos}, G.},
        title = "{Detection of a cyclotron line in SXP 15.3 during its 2017 outburst}",
      journal = {\mnras},
         year = 2018,
        month = oct,
       volume = {480},
       number = {1},
        pages = {L136-L140},
          doi = {10.1093/mnrasl/sly141},
archivePrefix = {arXiv},
       eprint = {1807.10696},
 primaryClass = {astro-ph.HE},
       adsurl = {https://ui.adsabs.harvard.edu/abs/2018MNRAS.480L.136M}
}

@ARTICLE{Galloway05,
       author = {{Galloway}, Duncan K. and {Wang}, Zhongxiang and {Morgan}, Edward H.},
        title = "{Discovery of Pulsations in the X-Ray Transient 4U 1901+03}",
      journal = {\apj},
         year = 2005,
        month = dec,
       volume = {635},
       number = {2},
        pages = {1217-1223},
          doi = {10.1086/497573},
archivePrefix = {arXiv},
       eprint = {astro-ph/0506247},
 primaryClass = {astro-ph},
       adsurl = {https://ui.adsabs.harvard.edu/abs/2005ApJ...635.1217G}
}

@ARTICLE{Mihara90,
       author = {{Mihara}, T. and {Makishima}, K. and {Ohashi}, T. and {Sakao}, T. and {Tashiro}, M.},
        title = "{New observations of the cyclotron absorption feature in Hercules X-1}",
      journal = {\nat},
         year = 1990,
        month = jul,
       volume = {346},
       number = {6281},
        pages = {250-252},
          doi = {10.1038/346250a0},
       adsurl = {https://ui.adsabs.harvard.edu/abs/1990Natur.346..250M}
}

@ARTICLE{Makishima90,
       author = {{Makishima}, K. and {Mihara}, T. and {Ishida}, M. and {Ohashi}, T. and {Sakao}, T. and {Tashiro}, M. and {Tsuru}, T. and {Kii}, T. and {Makino}, F. and {Murakami}, T. and {Nagase}, F. and {Tanaka}, Y. and {Kunieda}, H. and {Tawara}, Y. and {Kitamoto}, S. and {Miyamoto}, S. and {Yoshida}, A. and {Turner}, M.~J.~L.},
        title = "{Discovery of a Prominent Cyclotron Absorption Feature from the Transient X-Ray Pulsar X0331+53}",
      journal = {\apjl},
         year = 1990,
        month = dec,
       volume = {365},
        pages = {L59},
          doi = {10.1086/185888},
       adsurl = {https://ui.adsabs.harvard.edu/abs/1990ApJ...365L..59M}
}

@ARTICLE{Matsuoka09,
       author = {{Matsuoka}, Masaru and {Kawasaki}, Kazuyoshi and {Ueno}, Shiro and {Tomida}, Hiroshi and {Kohama}, Mitsuhiro and {Suzuki}, Motoko and {Adachi}, Yasuki and {Ishikawa}, Masaki and {Mihara}, Tatehiro and {Sugizaki}, Mutsumi and {Isobe}, Naoki and {Nakagawa}, Yujin and {Tsunemi}, Hiroshi and {Miyata}, Emi and {Kawai}, Nobuyuki and {Kataoka}, Jun and {Morii}, Mikio and {Yoshida}, Atsumasa and {Negoro}, Hitoshi and {Nakajima}, Motoki and {Ueda}, Yoshihiro and {Chujo}, Hirotaka and {Yamaoka}, Kazutaka and {Yamazaki}, Osamu and {Nakahira}, Satoshi and {You}, Tetsuya and {Ishiwata}, Ryoji and {Miyoshi}, Sho and {Eguchi}, Satoshi and {Hiroi}, Kazuo and {Katayama}, Haruyoshi and {Ebisawa}, Ken},
        title = "{The MAXI Mission on the ISS: Science and Instruments for Monitoring All-Sky X-Ray Images}",
      journal = {\pasj},
         year = 2009,
        month = oct,
       volume = {61},
        pages = {999},
          doi = {10.1093/pasj/61.5.999},
archivePrefix = {arXiv},
       eprint = {0906.0631},
 primaryClass = {astro-ph.IM},
       adsurl = {https://ui.adsabs.harvard.edu/abs/2009PASJ...61..999M}
}

@ARTICLE{Sugizaki17,
       author = {{Sugizaki}, Mutsumi and {Mihara}, Tatehiro and {Nakajima}, Motoki and {Makishima}, Kazuo},
        title = "{Correlation between the luminosity and spin-period changes during outbursts of 12 Be binary pulsars observed by the MAXI/GSC and the Fermi/GBM}",
      journal = {\pasj},
         year = 2017,
        month = dec,
       volume = {69},
       number = {6},
          eid = {100},
        pages = {100},
          doi = {10.1093/pasj/psx119},
archivePrefix = {arXiv},
       eprint = {1709.07579},
 primaryClass = {astro-ph.HE},
       adsurl = {https://ui.adsabs.harvard.edu/abs/2017PASJ...69..100S}
}

@ARTICLE{Ji20,
       author = {{Ji}, L. and {Ducci}, L. and {Santangelo}, A. and {Zhang}, S. and {Suleimanov}, V. and {Tsygankov}, S. and {Doroshenko}, V. and {Nabizadeh}, A. and {Zhang}, S.~N. and {Ge}, M.~Y. and {Tao}, L. and {Bu}, Q.~C. and {Qu}, J.~L. and {Lu}, F.~J. and {Chen}, L. and {Song}, L.~M. and {Li}, T.~P. and {Xu}, Y.~P. and {Cao}, X.~L. and {Chen}, Y. and {Liu}, C.~Z. and {Cai}, C. and {Chang}, Z. and {Chen}, G. and {Chen}, T.~X. and {Chen}, Y.~B. and {Chen}, Y.~P. and {Cui}, W. and {Cui}, W.~W. and {Deng}, J.~K. and {Dong}, Y.~W. and {Du}, Y.~Y. and {Fu}, M.~X. and {Gao}, G.~H. and {Gao}, H. and {Gao}, M. and {Gu}, Y.~D. and {Guan}, J. and {Guo}, C.~C. and {Han}, D.~W. and {Huang}, Y. and {Huo}, J. and {Jia}, S.~M. and {Jiang}, L.~H. and {Jiang}, W.~C. and {Jin}, J. and {Jin}, Y.~J. and {Kong}, L.~D. and {Li}, B. and {Li}, C.~K. and {Li}, G. and {Li}, M.~S. and {Li}, W. and {Li}, X. and {Li}, X.~B. and {Li}, X.~F. and {Li}, Y.~G. and {Li}, Z.~W. and {Liang}, X.~H. and {Liao}, J.~Y. and {Liu}, B.~S. and {Liu}, G.~Q. and {Liu}, H.~X. and {Liu}, H.~W. and {Liu}, X.~J. and {Liu}, Y.~N. and {Lu}, B. and {Lu}, X.~F. and {Luo}, Q. and {Luo}, T. and {Ma}, X. and {Meng}, B. and {Nang}, Y. and {Nie}, J.~Y. and {Ou}, G. and {Sai}, N. and {Shang}, R.~C. and {Song}, X.~Y. and {Sun}, L. and {Tan}, Y. and {Tuo}, Y.~L. and {Wang}, C. and {Wang}, G.~F. and {Wang}, J. and {Wang}, P.~J. and {Wang}, W.~S. and {Wang}, Y.~S. and {Wen}, X.~Y. and {Wu}, B.~Y. and {Wu}, B.~B. and {Wu}, M. and {Xiao}, G.~C. and {Xiao}, S. and {Xiong}, S.~L. and {Xu}, H. and {Yang}, J.~W. and {Yang}, S. and {Yang}, Yan-Ji and {Yang}, Yi-Jung and {Yi}, Q.~B. and {Yin}, Q.~Q. and {You}, Y. and {Zhang}, A.~M. and {Zhang}, C.~M. and {Zhang}, F. and {Zhang}, H.~M. and {Zhang}, J. and {Zhang}, P. and {Zhang}, T. and {Zhang}, W. and {Zhang}, W.~C. and {Zhang}, W.~Z. and {Zhang}, Yi and {Zhang}, Y.~F. and {Zhang}, Y.~J. and {Zhang}, Y.~H. and {Zhang}, Yue and {Zhang}, Zhao and {Zhang}, Zhi and {Zhang}, Z.~L. and {Zhao}, H.~S. and {Zhao}, X.~F. and {Zheng}, S.~J. and {Zhou}, D.~K. and {Zhou}, J.~F. and {Zhu}, Y.~X. and {Zhu}, Y. and {Zhuang}, R.~L.},
        title = "{Switches between accretion structures during flares in 4U 1901+03}",
      journal = {\mnras},
         year = 2020,
        month = apr,
       volume = {493},
       number = {4},
        pages = {5680-5692},
          doi = {10.1093/mnras/staa569},
archivePrefix = {arXiv},
       eprint = {2002.08919},
 primaryClass = {astro-ph.HE},
       adsurl = {https://ui.adsabs.harvard.edu/abs/2020MNRAS.493.5680J}
}

@ARTICLE{Apparao91,
       author = {{Apparao}, Krishna M.~V.},
        title = "{Flares in the X-Ray Source EXO 2030+375}",
      journal = {\apj},
         year = 1991,
        month = jul,
       volume = {375},
        pages = {701},
          doi = {10.1086/170235},
       adsurl = {https://ui.adsabs.harvard.edu/abs/1991ApJ...375..701A}
}

@ARTICLE{Cannizzo97,
       author = {{Cannizzo}, John K.},
        title = "{The Nature of the Fluctuations Preceding the Giant Bursts in the Bursting Pulsar GRO J1744-28}",
      journal = {\apj},
         year = 1997,
        month = jun,
       volume = {482},
       number = {1},
        pages = {178-181},
          doi = {10.1086/304150},
       adsurl = {https://ui.adsabs.harvard.edu/abs/1997ApJ...482..178C}
}

@ARTICLE{Doroshenko12,
       author = {{Doroshenko}, V. and {Santangelo}, A. and {Ducci}, L. and {Klochkov}, D.},
        title = "{Supergiant, fast, but not so transient 4U 1907+09}",
      journal = {\aap},
         year = 2012,
        month = dec,
       volume = {548},
          eid = {A19},
        pages = {A19},
          doi = {10.1051/0004-6361/201220085},
archivePrefix = {arXiv},
       eprint = {1210.4428},
 primaryClass = {astro-ph.HE},
       adsurl = {https://ui.adsabs.harvard.edu/abs/2012A&A...548A..19D}
}

@ARTICLE{Ducci19,
       author = {{Ducci}, Lorenzo and {Mereghetti}, Sandro and {Santangelo}, Andrea},
        title = "{Awakening of the Fast-spinning Accreting Be/X-Ray Pulsar A0538-66}",
      journal = {\apjl},
         year = 2019,
        month = aug,
       volume = {881},
       number = {1},
          eid = {L17},
        pages = {L17},
          doi = {10.3847/2041-8213/ab32f0},
archivePrefix = {arXiv},
       eprint = {1907.08078},
 primaryClass = {astro-ph.HE},
       adsurl = {https://ui.adsabs.harvard.edu/abs/2019ApJ...881L..17D}
}

@ARTICLE{Moon03,
       author = {{Moon}, Dae-Sik and {Eikenberry}, Stephen S. and {Wasserman}, Ira M.},
        title = "{SMC X-1 as an Intermediate-Stage Flaring X-Ray Pulsar}",
      journal = {\apjl},
         year = 2003,
        month = jan,
       volume = {582},
       number = {2},
        pages = {L91-L94},
          doi = {10.1086/367782},
archivePrefix = {arXiv},
       eprint = {astro-ph/0209414},
 primaryClass = {astro-ph},
       adsurl = {https://ui.adsabs.harvard.edu/abs/2003ApJ...582L..91M}
}

@ARTICLE{Postnov08,
       author = {{Postnov}, K. and {Staubert}, R. and {Santangelo}, A. and {Klochkov}, D. and {Kretschmar}, P. and {Caballero}, I.},
        title = "{The appearance of magnetospheric instability in flaring activity at the onset of X-ray outbursts in A0535+26}",
      journal = {\aap},
         year = 2008,
        month = mar,
       volume = {480},
       number = {2},
        pages = {L21-L24},
          doi = {10.1051/0004-6361:20079277},
archivePrefix = {arXiv},
       eprint = {0801.3165},
 primaryClass = {astro-ph},
       adsurl = {https://ui.adsabs.harvard.edu/abs/2008A&A...480L..21P}
}

@ARTICLE{Reig08-flare,
       author = {{Reig}, P. and {Belloni}, T. and {Israel}, G.~L. and {Campana}, S. and {Gehrels}, N. and {Homan}, J.},
        title = "{Bright flares from the X-ray pulsar SWIFT J1626.6-5156}",
      journal = {\aap},
         year = 2008,
        month = jul,
       volume = {485},
       number = {3},
        pages = {797-805},
          doi = {10.1051/0004-6361:200809457},
archivePrefix = {arXiv},
       eprint = {0804.3445},
 primaryClass = {astro-ph},
       adsurl = {https://ui.adsabs.harvard.edu/abs/2008A&A...485..797R}
}

@ARTICLE{Taam88,
       author = {{Taam}, Ronald E. and {Fryxell}, B.~A. and {Brown}, D.~A.},
        title = "{A Model for the Recurrent Flares in EXO 2030+375}",
      journal = {\apjl},
         year = 1988,
        month = aug,
       volume = {331},
        pages = {L117},
          doi = {10.1086/185248},
       adsurl = {https://ui.adsabs.harvard.edu/abs/1988ApJ...331L.117T}
}

@ARTICLE{Woods2000,
       author = {{Woods}, Peter M. and {Kouveliotou}, Chryssa and {van Paradijs}, Jan and {Koshut}, Thomas M. and {Finger}, Mark H. and {Briggs}, Michael S. and {Fishman}, Gerald J. and {Lewin}, W.~H.~G.},
        title = "{Detailed Analysis of the Pulsations during and after Bursts from the Bursting Pulsar (GRO J1744-28)}",
      journal = {\apj},
         year = 2000,
        month = sep,
       volume = {540},
       number = {2},
        pages = {1062-1068},
          doi = {10.1086/309367},
archivePrefix = {arXiv},
       eprint = {astro-ph/0004232},
 primaryClass = {astro-ph},
       adsurl = {https://ui.adsabs.harvard.edu/abs/2000ApJ...540.1062W}
}

@ARTICLE{Beri2015,
       author = {{Beri}, Aru and {Paul}, Biswajit and {Dewangan}, Gulab C.},
        title = "{Pulse-phase dependence of emission lines in the X-ray pulsar 4U 1626-67}",
      journal = {\mnras},
         year = 2015,
        month = jul,
       volume = {451},
       number = {1},
        pages = {508-516},
          doi = {10.1093/mnras/stv922},
archivePrefix = {arXiv},
       eprint = {1504.06225},
 primaryClass = {astro-ph.HE},
       adsurl = {https://ui.adsabs.harvard.edu/abs/2015MNRAS.451..508B}
}

@ARTICLE{DalFiume1998,
       author = {{dal Fiume}, D. and {Orlandini}, M. and {Cusumano}, G. and {del Sordo}, S. and {Feroci}, M. and {Frontera}, F. and {Oosterbroek}, T. and {Palazzi}, E. and {Parmar}, A.~N. and {Santangelo}, A. and {Segreto}, A.},
        title = "{The broad-band (0.1-200 keV) spectrum of HER X-1 observed with BeppoSAX}",
      journal = {\aap},
         year = 1998,
        month = jan,
       volume = {329},
        pages = {L41-L44},
          doi = {10.48550/arXiv.astro-ph/9711295},
archivePrefix = {arXiv},
       eprint = {astro-ph/9711295},
 primaryClass = {astro-ph},
       adsurl = {https://ui.adsabs.harvard.edu/abs/1998A&A...329L..41D}
}

@ARTICLE{Maitra13,
       author = {{Maitra}, Chandreyee and {Paul}, Biswajit},
        title = "{Pulse-phase-resolved Spectroscopy of Vela X-1 with Suzaku}",
      journal = {\apj},
         year = 2013,
        month = feb,
       volume = {763},
       number = {2},
          eid = {79},
        pages = {79},
          doi = {10.1088/0004-637X/763/2/79},
archivePrefix = {arXiv},
       eprint = {1212.1538},
 primaryClass = {astro-ph.HE},
       adsurl = {https://ui.adsabs.harvard.edu/abs/2013ApJ...763...79M}
}

@ARTICLE{Alonso-Hernandez22,
       author = {{Alonso-Hern{\'a}ndez}, J. and {F{\"u}rst}, F. and {Kretschmar}, P. and {Caballero}, I. and {Joyce}, A.~M.},
        title = "{Common patterns in pulse profiles of high-mass X-ray binaries}",
      journal = {\aap},
         year = 2022,
        month = jun,
       volume = {662},
          eid = {A62},
        pages = {A62},
          doi = {10.1051/0004-6361/202141774},
archivePrefix = {arXiv},
       eprint = {2201.09580},
 primaryClass = {astro-ph.HE},
       adsurl = {https://ui.adsabs.harvard.edu/abs/2022A&A...662A..62A}
}

@ARTICLE{Beri2014,
       author = {{Beri}, Aru and {Jain}, Chetana and {Paul}, Biswajit and {Raichur}, Harsha},
        title = "{Torque reversals and pulse profile of the pulsar 4U 1626-67}",
      journal = {\mnras},
         year = 2014,
        month = apr,
       volume = {439},
       number = {2},
        pages = {1940-1947},
          doi = {10.1093/mnras/stu087},
archivePrefix = {arXiv},
       eprint = {1401.2936},
 primaryClass = {astro-ph.HE},
       adsurl = {https://ui.adsabs.harvard.edu/abs/2014MNRAS.439.1940B}
}

@ARTICLE{Virtanen20,
       author = {{Virtanen}, Pauli and {Gommers}, Ralf and {Oliphant}, Travis E. and {Haberland}, Matt and {Reddy}, Tyler and {Cournapeau}, David and {Burovski}, Evgeni and {Peterson}, Pearu and {Weckesser}, Warren and {Bright}, Jonathan and {van der Walt}, St{\'e}fan J. and {Brett}, Matthew and {Wilson}, Joshua and {Millman}, K. Jarrod and {Mayorov}, Nikolay and {Nelson}, Andrew R.~J. and {Jones}, Eric and {Kern}, Robert and {Larson}, Eric and {Carey}, C.~J. and {Polat}, {\.I}lhan and {Feng}, Yu and {Moore}, Eric W. and {VanderPlas}, Jake and {Laxalde}, Denis and {Perktold}, Josef and {Cimrman}, Robert and {Henriksen}, Ian and {Quintero}, E.~A. and {Harris}, Charles R. and {Archibald}, Anne M. and {Ribeiro}, Ant{\^o}nio H. and {Pedregosa}, Fabian and {van Mulbregt}, Paul and {SciPy 1. 0 Contributors}},
        title = "{SciPy 1.0: fundamental algorithms for scientific computing in Python}",
      journal = {Nature Methods},
         year = 2020,
        month = feb,
       volume = {17},
        pages = {261-272},
          doi = {10.1038/s41592-019-0686-2},
archivePrefix = {arXiv},
       eprint = {1907.10121},
 primaryClass = {cs.MS},
       adsurl = {https://ui.adsabs.harvard.edu/abs/2020NatMe..17..261V}
}

@ARTICLE{Klochkov11,
       author = {{Klochkov}, D. and {Ferrigno}, C. and {Santangelo}, A. and {Staubert}, R. and {Kretschmar}, P. and {Caballero}, I. and {Postnov}, K. and {Wilson-Hodge}, C.~A.},
        title = "{Quasi-periodic flares in EXO 2030+375 observed with INTEGRAL}",
      journal = {\aap},
         year = 2011,
        month = dec,
       volume = {536},
          eid = {L8},
        pages = {L8},
          doi = {10.1051/0004-6361/201118185},
archivePrefix = {arXiv},
       eprint = {1111.3804},
 primaryClass = {astro-ph.HE},
       adsurl = {https://ui.adsabs.harvard.edu/abs/2011A&A...536L...8K}
}

@ARTICLE{Romanova08,
       author = {{Romanova}, Marina M. and {Kulkarni}, Akshay K. and {Lovelace}, Richard V.~E.},
        title = "{Unstable Disk Accretion onto Magnetized Stars: First Global Three-dimensional Magnetohydrodynamic Simulations}",
      journal = {\apjl},
         year = 2008,
        month = feb,
       volume = {673},
       number = {2},
        pages = {L171},
          doi = {10.1086/527298},
archivePrefix = {arXiv},
       eprint = {0711.0418},
 primaryClass = {astro-ph},
       adsurl = {https://ui.adsabs.harvard.edu/abs/2008ApJ...673L.171R}
}

@ARTICLE{Kulkarni08,
       author = {{Kulkarni}, A.~K. and {Romanova}, M.~M.},
        title = "{Accretion to magnetized stars through the Rayleigh-Taylor instability: global 3D simulations}",
      journal = {\mnras},
         year = 2008,
        month = may,
       volume = {386},
       number = {2},
        pages = {673-687},
          doi = {10.1111/j.1365-2966.2008.13094.x},
archivePrefix = {arXiv},
       eprint = {0802.1759},
 primaryClass = {astro-ph},
       adsurl = {https://ui.adsabs.harvard.edu/abs/2008MNRAS.386..673K}
}

\begin{appendix}
\section{Spin frequencies and energy-resolved profiles}
\begin{table}
    \centering
    \caption{The barycentre-corrected and orbital-corrected spin frequency measurements. All errors reported in this table are at 68\% (1$\sigma$) confidence level.}
    \resizebox{0.99\linewidth}{!}{
    \begin{tabular}{ccccc}
    \hline
OBS ID	&	Epoch (MJD)	&	$\nu_{\rm bary}$ (mHz) & $\nu_{\rm orb}$ (mHz)\\
\hline
\multicolumn{4}{c}{\nicer}\\
7204300101	&	60402.940 & 124.5816 $\pm$ 0.0043     & 124.5420 $\pm$ 0.0025	\\
7204300103	&	60404.645 & 124.5887 $\pm$ 0.0001     & 124.5484 $\pm$ 0.0005	\\
7204300104	&	60405.333 & 124.5878 $\pm$ 0.0008     & 124.5494 $\pm$ 0.0014	\\
7204300105	&	60406.393 & 124.5870 $\pm$ 0.0012     & 124.5539 $\pm$ 0.0008	\\
7204300106	&	60407.390 & 124.5809 $\pm$ 0.0005     & 124.5556 $\pm$ 0.0006	\\
7204300107	&	60408.300 & 124.5736 $\pm$ 0.0005     & 124.5572 $\pm$ 0.0005	\\
7204300110	&	60411.674 & 124.5445 $\pm$ 0.0019     & 124.5643 $\pm$ 0.0017	\\
7204300111	&	60412.291 & 124.5380 $\pm$ 0.0031     & 124.5640 $\pm$ 0.0037	\\
7204300115	&	60418.682 & 124.5437 $\pm$ 0.0004     & 124.5723 $\pm$ 0.0002	\\
\multicolumn{4}{c}{\astrosat}\\
9000006180 &	60411.000 & 124.54877 $\pm$ 0.00015 & 124.56286 $\pm$ 0.00015 \\
\multicolumn{4}{c}{\nustar}\\
91001317002	&	60413.000 & 124.5349 $\pm$ 0.0002     & 124.5665 $\pm$ 0.0002	\\
\hline  
\end{tabular}}    
\label{tab:nicerspin}
\end{table}

\begin{figure}
\centering
 \includegraphics[width=0.9\linewidth, height=10cm]{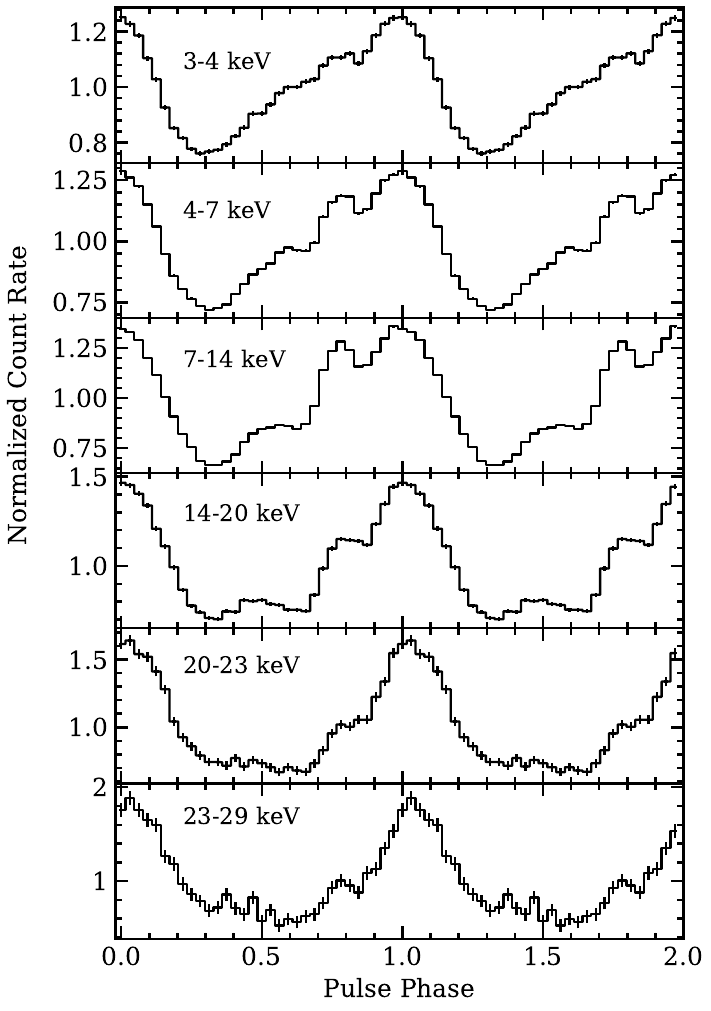}
 \caption{Comparison of pulse profiles in different energy bands: 3--4 keV, 4--7 keV, 7--14 keV, 14--20 keV, 20--23 keV and 23--25 keV.}
\label{fig:e-res-pp}
\end{figure}

\end{appendix}

\end{document}